\documentclass{ieee-ojvt}
\usepackage{cite}
\usepackage{amsmath,amssymb,amsfonts}
\usepackage{algorithmic}
\usepackage{algorithm}
\usepackage{graphicx,color}
\usepackage{textcomp}
\def\BibTeX{{\rm B\kern-.05em{\sc i\kern-.025em b}\kern-.08em
    T\kern-.1667em\lower.7ex\hbox{E}\kern-.125emX}}
\AtBeginDocument{\definecolor{ojcolor}{cmyk}{0.93,0.59,0.15,0.02}}

\begin{document}

\receiveddate{XX Month, XXXX}
\reviseddate{XX Month, XXXX}
\accepteddate{XX Month, XXXX}
\publisheddate{XX Month, XXXX}
\currentdate{19 August, 2024}

\title{Optimizing QoS in HD Map Updates: Cross-Layer Multi-Agent with Hierarchical and Independent Learning}

\author{Jeffrey Redondo$^1$ \IEEEmembership{(Student Member, IEEE)}, Nauman Aslam$^1$ \IEEEmembership{(Senior Member, IEEE)}, \\ Juan Zhang$^1$ \IEEEmembership{(Member, IEEE)}, Zhenhui Yuan$^2$ \IEEEmembership{(Member, IEEE)}}
\affil{University of Northumbria, Newcastle, UK}
\affil{University of Warwick, Coventry, UK}
\corresp{CORRESPONDING AUTHOR: J. Redondo (e-mail: jeffrey.redondo@northumbria.ac.uk).}
\markboth{Preparation of Papers for IEEE OPEN JOURNALS}{Author \textit{et al.}}

\begin{abstract}
The data collected by autonomous vehicle (AV) sensors such as LiDAR and cameras is crucial for creating high-definition (HD) maps to provide higher accuracy and enable a higher level of automation. Nevertheless, offloading this large volume of raw data to edge servers leads to increased latency due to network congestion in highly dense environments such as Vehicular Adhoc networks (VANET). To address this challenge, researchers have focused on the dynamic allocation of minimum contention window (CWmin) value. While this approach could be sufficient for fairness, it might not be adequate for prioritizing different services, as it also involves other parameters such as maximum contention window (CWmax) and infer-frame space number (IFSn).  In response to this, we extend the scope of previous solutions to include the control of not only CWmin but also the adjustment of two other parameters in the standard IEEE802.11: CWmax and IFSn, alongside waiting transmission time. To achieve this, we introduced a methodology involving a cross-layer solution between the application and MAC layers. Additionally, we utilised multi-agent techniques, emphasising a hierarchical structure and independent learning (IL) to improve latency to efficiently handle map updates while interacting with multiple services. This approach demonstrated an improvement in latency against the standard IEEE802.11p EDCA by  $31\%$, $49\%$, $87.3\%$, and $64\%$ for Voice, Video, HD Map, and Best-effort, respectively.
\end{abstract}

\begin{IEEEkeywords}
Edge computing, HD map, hierarchical learning, latency, multi-agent, offloading, reinforcement learning
\\	\fbox{\parbox{\dimexpr\columnwidth+20\fboxsep+30\fboxrule\relax}{
			This work has been submitted to the IEEE for possible publication. Copyright may be transferred without notice, after which this version may no longer be accessible.
	}}
\end{IEEEkeywords}

\maketitle

\section{Introduction}
HD maps hold the potential to elevate automation in autonomous vehicles (AVs) \cite{nvidea_hdmap}, playing a crucial role in realising the vision of widespread deployment on roads by 2025\cite{gov_uk,av_2025}. HD maps are engineered to deliver precise, centimetre-level information to AVs. This level of precision is imperative for ensuring the safe and efficient operation of AVs on public roads. Nonetheless, to produce an HD Map, the raw data generated by the AV's sensors must be processed, which increases the onboard-unit (OBU) load \cite{chameleon}. Offloading data to fog, edge, or cloud servers can mitigate the load increase on the OBU and reduce processing time. Previous research \cite{hdmap_processing_time} has reported that offloading data to edge servers could reduce processing by $66\%$. However, to guarantee a lower processing time, the wireless communication system must ensure the latency and throughput requirements in a dense scenario like VANET. Since the vehicular network widely utilises the standard IEEE802.11p, it faces the challenge of packet collision, as highlighted in \cite{ieee80211_cw} and \cite{ieee80211_cw2} for the high-dense environments such as VANET. These collisions translate into higher latency due to re-transmission. Several efforts have been made to reduce the latency caused by packet collision. For instance, in \cite{sojourn_time_handover}, authors have reduced latency by reducing the number of handovers, avoiding unnecessary overheads, and re-transmissions. Other studies such as \cite{avaq_edca_new_ac,performance_analysis_HDMAP,low_latency_new_ac} have introduced and implemented a new Access Category (AC) for low latency applications. In \cite{adaptive_edca}, the authors have applied game theory, while others have developed Q-learning algorithms to enhance fairness, as detailed in \cite{q_learning_fairness}. Additionally, some have chosen multiple access strategies, as discussed in \cite{TDMA} and \cite{HMAC}.
Nevertheless, solutions that involve a new AC also require an adaptable allocation value strategy for the CW parameters to provide Quality of Service (QoS) in dynamic networks. Solutions addressing the challenge of finding an optimal CW value as discussed in \cite{q_learning_fairness}, \cite{adaptive_edca}, and \cite{q_learning_edca_policy_RL} primarily focus on a single parameter, such as CW or CWmin. However, additional parameters, such as CWmax and IFSn, have yet to be evaluated and are also essential to improve the QoS. Accordingly, the optimization problem must focus on finding the optimal value for these parameters. However, developing a single-agent solution to allocate all three optimal values increases complexity by expanding the state and action space for the RL algorithm. To mitigate this problem, researchers have adopted the approach of dividing the main task into sub-tasks, demonstrating its effectiveness in \cite{multi_agent_edgemap}, and \cite{multi_agent_two_stage_DQN}. Their methodologies consist of subdividing the offloading decision and resource allocation as designed in \cite{multi_agent_edgemap}, as well as developing a multi-stage solution in \cite{multi_agent_two_stage_DQN} to improve the network performance. Subsequently, the question arises: How could incorporating new parameters such as IFSn and transmission time, along with employing multi-agents, each assigned a unique task, enhance network performance?

In this paper, we address the aforementioned question by adopting a task subdivision approach. These subtasks involve finding the optimal CWmin and CWmax pair, determining the IFSn, and allocating the waiting transmission time. We have integrated a hierarchical and individual machine learning (ML) technique, assigning one task to each agent. In addition, we have developed a cross-layer design permitting seamless communication between the application layer and the MAC layer to access the actions provided by the agents in regard to CWmin, CWmax, and IFS values. Our study demonstrated the improvement in lowering the latency by breaking the problem into sub-tasks. Moreover, we assessed the impact of the dissemination of different types of data (Voice (VO), Video (VI), and Best-effort (BE)) simultaneously with HD Map data type, which previous studies have neglected. Therefore, there is an opportunity to continue enhancing the wireless network with a new ML using a hierarchical and individual methodology to overcome the problems above in disseminating HD maps and other types of services in a vehicular network.

\subsection{Challenges} 
In wireless communication systems based on the IEEE802.11 standard, a persistent challenge revolves around packet collisions, particularly in highly dense environments, as highlighted in studies such as \cite{ieee80211_cw} and \cite{ieee80211_cw2}. This challenge is amplified in VANETs due to the uncertainties in traffic flow, which introduce factors such as high mobility and density. To address this, researchers have suggested several solutions to reduce packet collision by utilising new strategies to improve the EDCA by implementing new AC \cite{low_latency_new_ac} or using RL \cite{q_learning_edca_policy_RL} to improve the latency of the network. Nevertheless, their implementation implies modification to the existing standard, which is undesirable. Subsequently, we propose an RL-based cross-layer solution that seamlessly communicates between the application and MAC layer. Implementing a solution that leverages RL, which involves finding the optimal value of the three parameters, CWmin-max pair, IFSn, and waiting transmission time, will inevitably expand the state and action space as the number of vehicles increases. As noted in \cite{hierachical_rl_book_chap}, the number of state and action variables grows the complexity of the problem representation exponentially. Humans have exploited hierarchical organizations to tackle highly complex tasks. Thus, it is opportunely to adopt a hierarchical approach to solve this complex problem of finding the optimal value of the three parameters mentioned above for our solution. The authors in \cite{hierarchical_RL_survey} also noted that learning a subtask can be easier. A similar approach to hierarchical learning is the divide-and-conquer strategy. Subsequently, we designed an RL algorithm consisting of three agents, each solving a specific task, effectively leveraging task subdivision and hierarchical principles. We distinguish from conventional hierarchical models by adopting a strategy in which agents operate simultaneously within the environment rather than sequentially. This approach combines hierarchical and individual learner strategies. Specifically, There are two agents which interact in a hierarchical approach where the principal agent alters the state space of the secondary agent, directly affecting the secondary agent’s actions. In contrast, a third agent, which functions as an individual learner, acts independently of the first two agents. Interaction between the hierarchical agents and the individual learner is facilitated through a shared reward function and environmental observations. This design choice eliminates the need for direct communication among agents and reduces the latency associated with waiting for instructions, thereby streamlining the decision-making process and enhancing overall efficiency. Our contributions are as follows:

\subsection{Contributions} 
\begin{itemize}
\item \textbf{A Optimal Solution for QoS Requirements in Multi-Service Environments:} We addressed how multi-agent and task subdivisions could aid in finding the optimal solution to fulfil the QoS requirement for the dissemination of HD maps in a multi-service environment by designing and successfully integrating a hierarchical and individual learning RL algorithm.
\item \textbf{A Novel Cross-Layer Solution for IFSn and CWmin Optimization}: We have developed a novel cross-layer solution between the application and MAC layers, which dynamically allocates both the CW and IFSn values through two agents in a hierarchical structure. This approach leads to a significant reduction in latency in vehicular communication networks.
\item \textbf{Enhanced Cross-Layer Design with Individual Learner for Application Layer}: We expanded the cross-layer design by integrating a third agent. This agent is considered an individual learner and functions at the application layer, avoiding involvement in the standard IEEE802.11p MAC layer. The solution involves integrating three agents, each playing a distinct role. The third agent independently focuses on learning the most efficient transmission waiting time to control when the vehicles will transmit. This subdivision of tasks into three has optimized data dissemination, particularly in scenarios involving HD map data showing a substantial decrease in service latency.
\end{itemize}

The rest of the paper is organized as follows: Section \ref{related_work} includes a comprehensive literature review on CW, EDCA, and multi-agent ML topics. Section \ref{problem_statement} describes the problem statement, while Section \ref{design} explains the design of the Q-learning algorithm. Section \ref{simulation} introduces the simulation setup used in our experiments. Section \ref{results} presents the results and discussion. Finally, the conclusions are presented in section \ref{conclusion}.
\section{Related Work} \label{related_work}
Several investigations have been conducted to improve IEEE802.11. These studies focus on implementing new access categories and allocating CW value. The latest research endeavour has focused on implementing single-agent or multi-agent methodologies to find hidden patterns in dense scenarios.

\subsection{Contention Windows \& EDCA}
One of the first approaches to optimise the latency for wireless networks based on IEEE802.11 technology is optimising queue mapping by applying different priorities. For instance, in \cite{q_learning_edca_policy_RL}, authors focused on mapping the packets by low and high AC priorities. Other researchers explored queue load to allocate the packet to an AC accordingly, as outlined in \cite{dynamic_queue}. However, these methods are insufficient as different or similar priorities might be allocated to other applications, generating more load to one AC. Therefore, authors have delved into the addition of new ACs and modifications to the EDCA to improve the delivery of data for low-latency application requirements, as seen in studies such as \cite{queue_IoV,logical_EDCA,adaptive_edca}. Even though these efforts have demonstrated improvement, there is still the fixed CW parameter value, as mentioned in \cite{ieee80211_cw} and \cite{ieee80211_cw2}, which continues to impact the network latency and throughput in dense scenarios. To address this challenge, researchers have developed solutions that utilise ML to learn the complex patterns of the mobile network and determine the optimal CW value dynamically. For instance, in \cite{q_learning_fairness}, authors introduced a single-agent RL algorithm that learns to identify the optimal CW value to ensure fairness among the vehicles and applications, resulting in approximately $12\%$ improvement. 
Nevertheless, the HD map data type, crucial for enabling a higher level of automation for AVs, has not been considered in the data dissemination methods mentioned in the abovementioned investigations. This results in categorizing the HD map data as BE, which has one of the lowest priorities impacting the overall average latency and throughput requirements to guarantee QoS. Furthermore, the parameters CWmax and IFSn are not considered in their optimisation problem, which is crucial for the avoidance of collisions and priority allocation between services. Therefore, we introduce a new solution to optimise three variables: CWminmax, IFSn, and waiting transmission time.

\subsection{Multi-agent Subdivision of Tasks}
One of the emerging approaches for improving network performance in wireless technology is the application of hierarchical RL (HRL) due to the subdivision of tasks. This approach shows promise in addressing the complexities of solving large tasks by subdividing the tasks into subtasks, which could facilitate learning the optimal policy \cite{hierarchical_RL_survey}. Furthermore, it has been proven that this approach outperforms the standard RL approach \cite{hierarchical_RL_survey}. This approach is also beneficial because the problem representation grows exponentially with the increase in the state and action space as stated in \cite{hierachical_rl_book_chap}. In implementing task subdivisions in wireless systems, initial investigations started focusing on splitting the offloading process and the resource allocation into subtasks instead of having an overall problem to solve. This methodology has resulted in an improvement in the overall network performance in terms of latency. For instance, the study on latency and energy awareness in MC-NOMA multi-access \cite{multi_agent_two_stage_DQN} has showcased the benefits of subdividing the offloading decision and the channel allocation in sub-tasks, each managed by an agent in hierarchical mode. The first agent takes the initial action, and the second agent concludes the final action after receiving the other's instruction. Another study \cite{multi_agent_edgemap} also has demonstrated the effectiveness of subdividing a single task into two subtasks to decrease the overall processing time for disseminating HD map updates. One subtask is related to offloading decisions, while the other relates to resource allocation. Another similar work has been conducted in \cite{multi_agent_wifi_5G} involving both offloading tasks and resource management in Wi-Fi and 5G, highlighting again the potential of task subdivision to simplify the learning process by allocating specific agents to handle each subprocess. 

However, it is essential to note that these solutions have used channel and resource allocation without considering the CW and IFSn parameters, and the coverage time variable within the same optimisation problem to minimize the overall average latency of the system per service.
Furthermore, they do not evaluate wireless systems that simultaneously disseminate multiple types of services. Therefore, we focus on developing a solution that adjusts to a high-density, high-mobility, and multi-service environment while finding the optimal CW and IFS values and waiting transmission time. To achieve optimal value allocation in a dynamic network, we integrated an HRL strategy and subdivided tasks. We established a hierarchical structure among the agent CW and IFS, where agent CW instructs with its action the agent IFS to take action accordingly. One difference from the previous solution is that both agents perform actions on the environment. Additionally, we added a third agent in IL mode, which focuses on allocating each vehicle's most suitable waiting transmission time.

\section{Problem Statement}\label{problem_statement}
To define our scenario, we begin by defining a set of categories (type of service) as $\mathcal{D} = \{d_{i},d_{i+1},...,d_M\}$, where $M \in \mathbb{N}^+$ and represents the total number of data categories and each index $i$ corresponds to a natural number such that $i \in \mathbb{N}^+$ and $i \leq M$. Each category $d_i$ represents a specific data type that an AV can transmit. The set of AVs is denoted by $\mathcal{C}=\{c_{j},c_{j+1},…,c_N\}$, where $N$ represents the total number of AVs in the system. Each AV $c_j$ is randomly assigned to one of the data categories $d_i$ and follows a specific route that defines its traffic flow. Additionally, we consider a discrete set of time slots $\mathcal{T} = \{t_{k}, t_{k+1}, \ldots, t_T\}$, with each time slot indexed by $k, T \in \mathbb{N}^+$ and $k \leq T$. These time slots model the time-dependent aspects of transmission and scheduling. Subsequently, the utility function for optimisation is inherently multi-objective as it seeks to balance two conflicting objectives: maximizing throughput and minimizing latency. To this end, we adopted the objective function as in \cite{adaptive_edca} because it encapsulates both throughput ($\mathcal{R}$) and latency ($\mathcal{L}$) over the time slots in $\mathcal{T}$. For enhancement, we included penalties and bonuses as $\mathcal{V}$ to provide the correct priority for each data category. Hence, the utility function is expressed as follows:
\vspace{0.03cm}
\begin{equation}
\begin{aligned}
    U(d(c_j), t) = \alpha_1 \frac{\mathcal{R}_{c}(d(c_j), t)}{\mathcal{R}_{\text{max}}(d(c_j), t)} - \alpha_2 \frac{\mathcal{L}_{c}(d(c_j), t)}{\mathcal{L}_{\text{max}}(d(c_j), t)} \\
    + \mathcal{V}_{c}(d(c_j), t), \quad
    \forall c \in \mathcal{C}, \forall d \in \mathcal{D}, \forall t \in \mathcal{T}
\end{aligned}
\label{eq:utility_function_penalty}
\end{equation}

The coefficients $\alpha_1$, and $\alpha_2$ are weights to provide a trade-off between $\mathcal{R}$ and $\mathcal{L}$. $\mathcal{R}_{max}$ refers to the maximum data rate established for the specific category $d$, and $\mathcal{L}_{max}$ corresponds to the maximum latency.

\begin{equation}
\begin{aligned}
    \mathcal{V}_{c}(d(c_j), t) = & \left( -b \cdot \mathbb{I}\left(\mathcal{R}_{c}(d(c_j), t) < \mathcal{R}_{\text{th}}\right) \right. \\
    & \left. + a \cdot \mathbb{I}\left(\mathcal{R}_{c}(d(c_j), t) > \mathcal{R}_{\text{th}}\right) \right) \\
    & + \left( a \cdot \mathbb{I}\left(\mathcal{L}_{c}(d(c_j), t) < \mathcal{L}_{\text{th}}\right) \right. \\
    & \left. - b \cdot \mathbb{I}\left(\mathcal{L}_{c}(d(c_j), t) > \mathcal{L}_{\text{th}}\right) \right)
\end{aligned}
\label{eq:penalty_v}
\end{equation}

The calculation of $\mathcal{V}$ is performed by the equation \eqref{eq:penalty_v}. The equation compared the current data rate with a data rate threshold $\mathcal{R}_{th}$, and the current latency with the latency threshold $\mathcal{L}_{th}$, the indicator function $\mathbb{I}$ takes value of $1$ when the condition is true, and $0$ when it is false. The value $a$ represents the bonus value and $b$ the penalty value.
\\
\\
The maximization problem is defined as follows:
\vspace{0.03cm}
\begin{equation}
    \underset{cw, ifs, wt}{\max} \sum_{c \in \mathcal{C}} \sum_{d(c_j) \in \mathcal{D}} \sum_{t \in \mathcal{T}} y_{c} U(d(c_j), t)
    \label{eq:max_problem}
\end{equation}

Subject to:
\vspace{0.03cm}
\begin{equation}
    y_c(t) = \mathbb{I}(wt - (t - t_0) = 0), \quad  wt,t,t_{0} \in \mathbb{R} 
    \label{eq:constraint_x}
\end{equation}
\begin{equation}
    \frac{1}{|\mathcal{C}|} \sum_{c=1}^{|\mathcal{C}|} \mathcal{L}_{c}(d(c_j), t) \leq \mathcal{L}_{\text{max}}(d(c_j), t), \quad \mathcal{L} \in \mathbb{R}
    \label{eq:constraint_L}
\end{equation}
\begin{equation}
    \sum_{c=1}^{|\mathcal{C}|} \mathcal{R}_{c}(d(c_j), t) \geq \mathcal{R}_{\text{min}}(d(c_j), t) , \quad  \mathcal{R} \in \mathbb{R}
    \label{eq:constraint_R}
\end{equation} 
\vspace{0.1cm}
\begin{equation}
    wt(d(c_j), t) \leq wt_{\text{max}}(d(c_j), t),\quad  wt, wt_{\text{max}} \in \mathbb{R}  ,\quad wt > 0
    \label{eq:constraint_w}
\end{equation}
\vspace{0.1cm}
\begin{equation}
\begin{aligned}
 CW_{\text{min}}(d(c_j), t) \leq CW_{\text{max}}(d(c_j), t), \quad CW_{\text{min}}, CW_{\text{max}} \in \mathbb{N}^+ , \\ \quad CW_{\text{min}}, CW_{\text{max}} >0
\end{aligned}
\label{eq:constraint_cw}
\end{equation}
\vspace{0.1cm}
\begin{equation}
\begin{aligned}
    IFSn(d(c_j)) \leq IFSn_{max}(d(c_j)) , \quad IFSn,IFSn_{max} \in \mathbb{N}^+ \\,\quad IFSn,IFSn_{\text{max}}  > 0
    \end{aligned}
\label{eq:constraint_ifs}
\end{equation}
As $y_{c}$ indicates whether the AV transmits or not, it is defined as a binary control variable. The problem is defined as mixed-integer linear programming because it involves a linear combination of continuous and integer/binary variables. Additionally, the variable $y_{c}$ is dependent on the waiting time $wt$, where $(t - t_0)$ is the elapsed time since the $wt$ was initialized at $t_0$. It measures how much time has passed since the counter started. The constraint \eqref{eq:constraint_L} refers to the maximum tolerable latency $\mathcal{L}_{\text{max}}$, \eqref{eq:constraint_R} is the minimum data rate $\mathcal{R}_{\text{min}}$, \eqref{eq:constraint_w} is the maximum waiting time $wt$, \eqref{eq:constraint_cw} is the maximum $CW$ allowed, and \eqref{eq:constraint_ifs} is the maximum $IFSn$ permitted. 

The optimisation problem is characterized by mixed-integer variables, a dynamic environment with high dimensionality, and the need for real-time decisions. This indicates a high level of complexity, suggesting that traditional optimization techniques may not be suitable for finding the optimal. Therefore, we adopted an RL paradigm that offers an alternative capable of learning complex patterns and adapting to a dynamic environment such as VANET. The algorithm will optimise the transmission decision $y_c$ in a way that maximizes system utility while respecting the constrains \eqref{eq:constraint_L}, \eqref{eq:constraint_R}, \eqref{eq:constraint_w}, \eqref{eq:constraint_cw}, and \eqref{eq:constraint_ifs}.   

\section{Design} \label{design}
This section illustrates a description of the hierarchical and individual learning design.

\subsection{Solution Architecture}
\begin{figure}[h]
  \begin{center}
  \includegraphics[width=\linewidth]{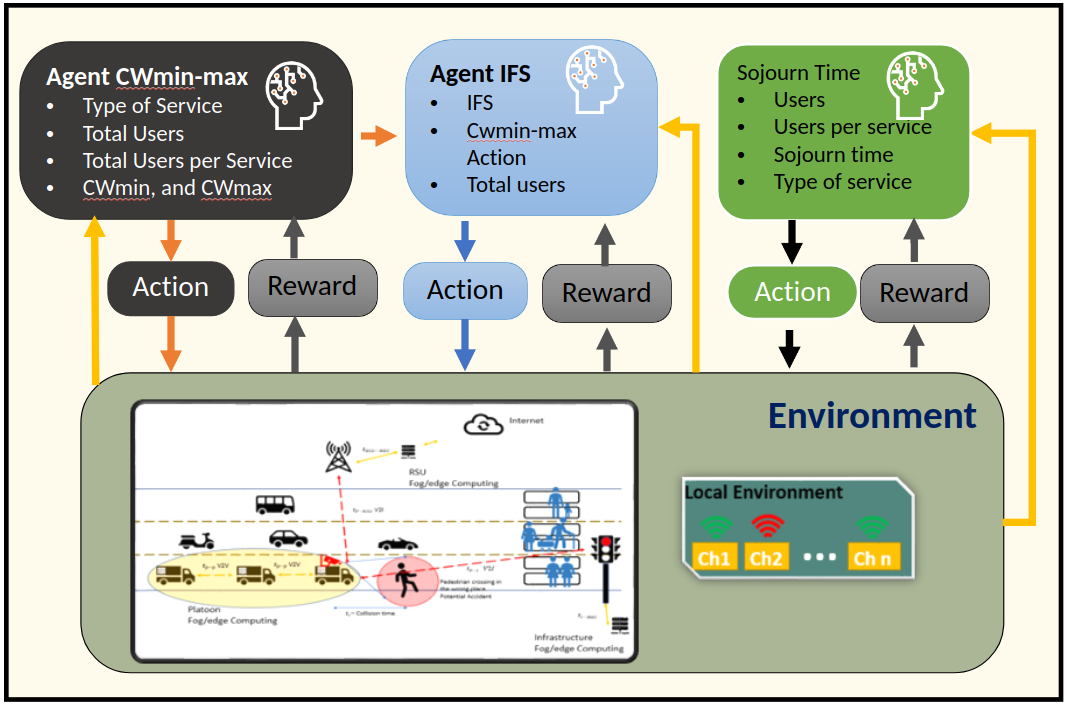}
  \caption{Multi-agent diagram.}
  \label{fig:crosslayer_design_3_agents}
  \end{center}
\end{figure}

Figure \ref{fig:crosslayer_design_3_agents} displays the RL learning scheme for three agents. The three agents are named as follows: Agent One is "Agent CWmin" " Agent Two is "Agent IFS", and Agent Three is "Agent wt". This design is characterized by the strategy applied to hierarchical and individual learning. The hierarchical approach refers to dividing a main task into smaller tasks. Thus, we established that allocating the wireless resources to achieve latency and throughput requirements is the main task. Even though this can be performed by using the current standard EDCA or its modification \cite{adaptive_edca,avaq_edca_new_ac, performance_analysis_HDMAP}, queue management \cite{dynamic_queue}, or utilizing an algorithm to allocate the suitable CW \cite{q_learning_fairness}. None of these solutions includes multi-service with HD map data dissemination or investigation of the allocation of CWmax and IFSn. Consequently, We have subdivided the main tasks into the following subtasks to achieve latency and throughput requirements. Subtask one is assigned to Agent CWmin to find the optimal CWminmax pair value; subtask two is assigned to Agent IFS to find the optimal IFSn value; and subtask three is assigned to Agent wt to find the most suitable waiting transmission time $wt$. 

In addition, we established hierarchical decision-making between agent CWmin and agent IFS; this is illustrated in Figure \ref{fig:crosslayer_design_3_agents}. Agent CWmin directly affects the decision-making of agent IFS by providing its action into the state space. This means that Agent IFS's immediate action is correlated with Agent CWmin's action. The rationale behind this is that both CW and IFSn impact channel allocation, so we combine these values and add a correlation. For independent learning, specify that all agents can act on the environment simultaneously. Furthermore, the third does not directly communicate with the other agents. It solely relies on its environment observation for feedback from the other agents, without any direct knowledge of their actions that could influence its decision-making.

\subsection{Cross-layer design}
\begin{figure}[h]
  \begin{center}
  \includegraphics[width=\linewidth]{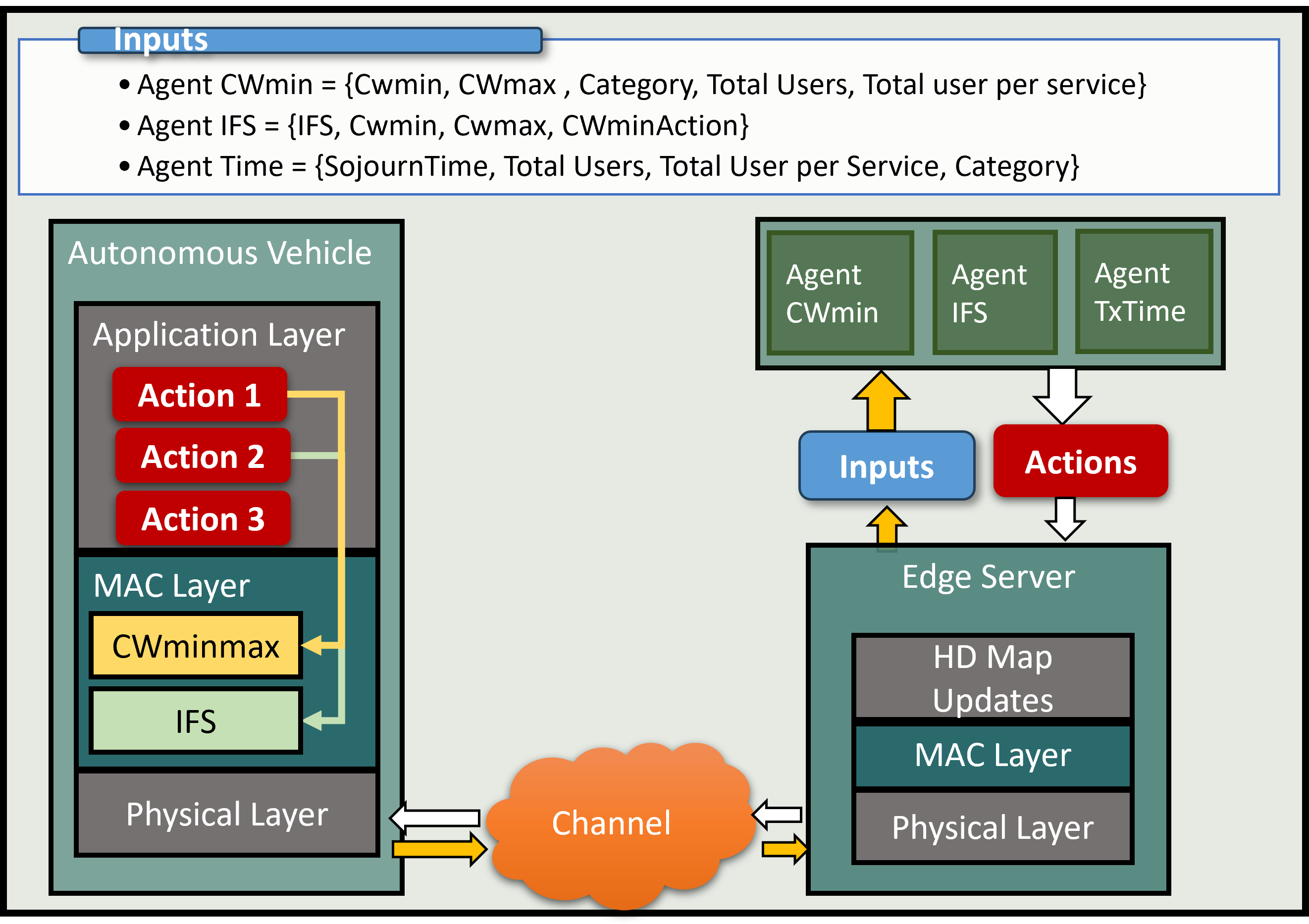}
  \caption{Cross-layer diagram application and MAC layer.}
  \label{fig:crosslayer_design_ifs_cwminmax_stime}
  \end{center}
\end{figure}

Figure \ref{fig:crosslayer_design_ifs_cwminmax_stime} illustrates the cross-layer architecture for the AV for the reception of the agent's actions to allocate the values CWminmax pair, IFSn and $wt$. As depicted in the figure on the left side, the design demonstrates a seamless communication link between the application and MAC layers. The right side of the figure corresponds to the edge server, where the agents operate within the application layer.
The agents' actions are transmitted to the respective AVs. Upon receiving the inputs and the agent's responses, the AVs process the packet and extract the actions directly at the application layer. For actions one and two, the cross-layer design allows the MAC layer to directly access the CWmin and CWmax values (action 1) and IFSn (action 2) without additional communication exchanges. The third action, $wt$, is utilised to control the timing of subsequent transmissions, and it is only part of the application layer.

The inputs for each agent are presented in the upper part of Figure \ref{fig:crosslayer_design_ifs_cwminmax_stime}. The inputs refer to the state space, which is further elaborated in the following subsection. The idea is that the agent must know the current environment status in which they are taking action. Therefore, part of its states contains the current value of the parameters they are changing, such as CWmin, CWmax, and IFSn. Additionally, Agent CWmin and Agent wt shared some features: Total number of users, total number of users per service (category), and Category).

\subsection{Reinforcement Learning}
Using deep neural neurons in RL could increase the power consumption during training as demonstrated in \cite{power_neural_neuron}. The study revealed that tuning of hyper-parameters, particularly batch size and learning rate, can significantly impact power consumption during training. In addition, in our previous work in \cite{our_sojourn_multi_agent}, we have tried  Therefore, as we aim to develop a lower computational capacity RL solution, we avoid using the neural network RL algorithm. Consequently, our algorithm design is based on Q-learning with temporal difference (TD). We use TD learning because it combines the strengths of Dynamic Programming (DP) and Monte Carlo (MC) methods. DT updates estimates after each step, similar to DP, for faster learning. Like MC, it operates without needing a complete model of the environment, making it effective in complex or unknown settings. In our RL design, the environment involving AVs is stochastic because of the vehicular network's high mobility and high density. Thus, the probability transition $P(s, s^\text{'})$ from the current state $s$ to the next $s^\text{'}$ is unknown. Since we don't have explicit knowledge of the transition probabilities, we cannot directly apply methods that rely on a model of the environment, such as DP. Therefore, we utilised Q-learning TD, which combines DP and MC methods. It is a model-free RL algorithm, meaning it does not require knowledge of the transition probabilities $P(s, s^\text{'})$. Instead, it learns an optimal policy $\pi^*$ by interacting with the environment and updating Q-values based on the observed rewards and transitions.

At time $t$, the agents observe the environment and select an action defined by $\pi(a,s)$ using an $\epsilon$-greedy strategy. After taking an action, the agents receive a reward. In our solution, the reward function is the same for all Agents, though their state and action spaces differ. Finally, the Q-value table is updated using the Q-learning update rule in equation \eqref{eq:q_learning}, where $\gamma$ (gamma) and $\alpha$ (alpha) represent the discount factor and learning rate, respectively.

\vspace{0.01cm}
\begin{equation}
    Q(s,a) = Q(s,a) + \alpha\left[r+\gamma*max_{a'}Q(s',a')-Q(s,a)\right]
\label{eq:q_learning}
\end{equation}
 Firstly, for simplicity, we denote $y = \gamma*max_{a'}Q(s',a')-Q(s,a)$. The utility function \eqref{eq:utility_function_penalty} is the reward function. Thus, by substituting \eqref{eq:utility_function_penalty} in \eqref{eq:q_learning}, the preceding equation can be expressed as,
 \vspace{0.01cm}
\begin{equation}
    Q(s,a) = Q(s,a) + \alpha\left[U+y\right]
\label{eq:q_learning_utility}
\end{equation}

\paragraph{\textbf{Hierarchical Structure}}
Our main distinction with the traditional hierarchical approach is having each subtask learned for an agent who takes actions directly to the environment independently. However, we leveraged the main advantage of the hierarchical approach by divide and conquer strategy \cite{hierachical_rl_book_chap}. Instead of having a single agent handle all actions, which could lead to exponential growth in the state and action space \cite{hierachical_rl_book_chap}, we distribute the tasks among multiple agents. Thus, firstly, we subdivide the task as described previously into three parts: one to find the CWminmax pair, second to find IFSn and third to allocate the waiting transmission time $wt$. Furthermore, we have created a hierarchical structure between Agent CW and Agent IFS by adding Agent CW's action into the state space of Agent IFS. This approach provides immediate communication between them, and Agent IFS needs to consider the other agent's instructions while taking action. This creates a correlation between their action, which is desirable because both parameters affect the network performance and QoS.
The agent CWmin learns the policy $\pi^{\text{cw}}$ as follows:
\vspace{0.01cm}
\begin{equation}
\pi^\text{cw} = \arg\max_{a_{\text{cw}}} Q(s_{\text{cw}}, a_{\text{cw}})
\label{eq:policy_agentcw}
\end{equation}
and the state space is,
\vspace{0.01cm}
\begin{equation}
S_\text{cw} = \{T_\text{v}, D, T_\text{cv}, CW_\text{min}, CW_\text{max}\}
\label{eq:state_agentcw}
\end{equation}
where $T_\text{v}$ corresponds to the total number of active vehicles, $T_\text{cv}$ is the total number of actives vehicles per category $d$. and $CW_\text{min}$ is the CW minimum value, and $CW_\text{max}$ is the CW maximum. Thus, the Q table value of the agent CW is updated as follows:
\vspace{0.01cm}
\begin{equation}
    Q_{CW_\text{min}}(s_\text{cw},a_\text{cw}) = Q(s_\text{cw},a) + \alpha\left[U+y\right]
\label{eq:q_learning_cw}
\end{equation}
The second agent IFS learns a policy 
\begin{equation}
\pi^\text{ifs} = \arg\max_{a_\text{ifs}} Q(s_\text{ifs}, a_\text{ifs})
\label{eq:policy_agentifs}
\end{equation}
and state space is,
\vspace{0.01cm}
\begin{equation}
S_\text{IFS} = \{CW_\text{min},CW_\text{max},\mathcal{A}_\text{cw}, \text{IFSn}\}
\label{eq:state_agentifs}
\end{equation}
where $\mathcal{A}_\text{cw}$ denotes the action of the agent CW. Then, the Q table value for the agent IFS is updated as follows,
\vspace{0.01cm}
\begin{equation}
    Q_\text{IFS}(s_\text{ifs},a_\text{ifs}) = Q(s_\text{ifs},a_\text{ifs}) + \alpha\left[U+y\right]
\label{eq:q_learning_ifs}
\end{equation}

\paragraph{\textbf{Independent Structure}}
The notion of independent strategy pertains to the practice of enabling individual agents to operate independently within the environment. Moreover, the third agent, denoted as Agent wt, possesses a unique characteristic: it does not communicate with the other two agents. Instead, it assimilates knowledge solely from the environment and the reward function. The reward function delineates network performance with respect to latency and throughput, as expressed in equation \ref{eq:utility_function_penalty}.
Thus, we defined the Agent wt as the independent agent. Its policy is defined as,
\vspace{0.01cm}
\begin{equation}
\pi^\text{wt} = \arg\max_{a_{\text{wt}}} Q(s_{\text{wt}}, a_{\text{wt}})
\label{eq:policy_agentwt}
\end{equation}
where the subscript wt refers to waiting time. For the state we have, 
\vspace{0.01cm}
\begin{equation}
    S_\text{wt} = \{Sj,T_\text{v}, C, T_\text{cv}\}
    \label{eq:state_agentwt}
\end{equation}
In a previous study (refer to \cite{our_sojourn_single_agent}), the state space was formally established, with $S_j$ denoting the sojourn time as outlined in the aforementioned source. This time variable represents the duration for which the vehicle remains within the coverage area of the base station. It is determined by considering the coverage area of the base station in relation to the distance and velocity of the AV. Consequently, the Q table value can be computed utilizing the following equation:
\begin{equation}
    Q_\text{wt}(s_\text{wt},a_\text{wt}) = Q(s_\text{wt},a_\text{wt}) + \alpha\left[U+y\right]
\label{eq:q_learning_wt}
\end{equation}

\subsection{Action and Process}
The standard IEEE802.11p has the parameter CW, which varies depending on encountering a collision because the channel is busy. When a collision occurs, the contention windows increase; this event corresponds to the backoff process. The transmission of the data then will start only when the backoff process is completed and the channel is idle, more details in \cite{q_learning_fairness, performance_analysis_HDMAP}. Additionally, if the EDCA is enabled, the CW values differ for each type of service. Therefore, we are focusing on dynamically adapting the CW minimum and maximum to deliver HD map data properly in a multi-service scenario. Our Algorithm \ref{algo:q_learning} considered three actions: decrease, maintain, and increase the CW minimum and maximum pair value. We have followed two approaches from the literature and the current standard to compare our design.
Firstly, We considered the action space in \cite{RL_cw_simple}, where the authors proposed a set of actions $A_1 = \{16, 32, 64, 128, 256, 512, 1024\}$ that we adapted to eight values $A'_1 = \{7, 16, 32, 64, 128, 256, 512, 1000\}$. For the second approach in \cite{q_learning_fairness}, we selected the set of Actions $A_2 = \{\frac{CW_t-1}{2}, CW_t, (CW*2-1)\}$, in this approach, the CW value is selected as per standard and then it dynamically changes as per the agent's action. As it is observed, both approaches do not have the CW maximum parameter, which is also important to consider to reduce latency and provide a priority level; this is also present in the EDCA, where each service has a different CWmin and CWmax. Finally, we considered the three actions in \cite{q_learning_fairness} with the difference that we introduced a set of CW minimum and maximum values per category to improve the QoS as follows:
\vspace{0.01cm}
\begin{equation}
CW_\text{min},CW_\text{max} =
\begin{cases}
    2,10 \quad \text{Voice Category} \\ 
    3,17 \quad \text{Video Category}\\
    3,17 \quad \text{HD Map Category}\\
    7,23 \quad \text{Best-effort Category}\\
\end{cases}
\label{eq:values_cw_min_max}
\end{equation}

\begin{figure}[h]
  \begin{center}
  \includegraphics[width=\linewidth]{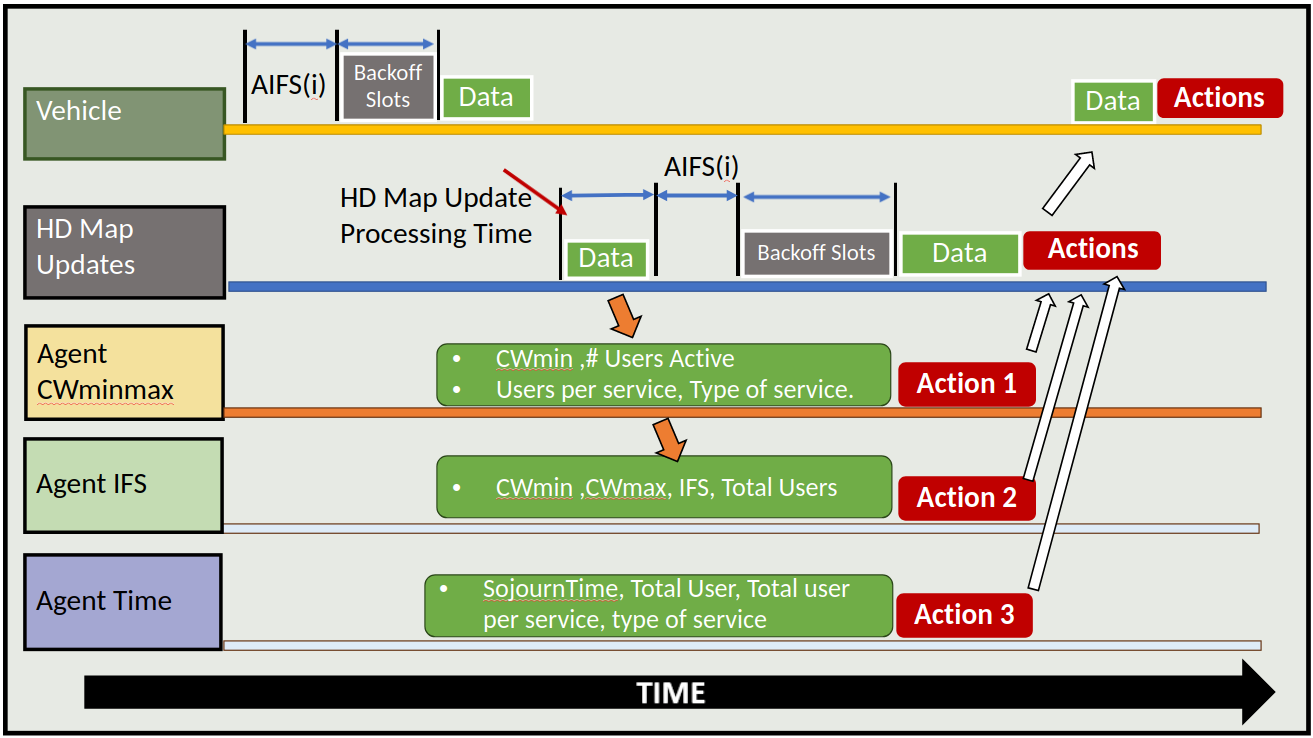}
  \caption{Data Flow, in the time domain, from Vehicle to Agent to Vehicle.}
  \label{fig:TimeFlow_3_agent}
  \end{center}
\end{figure}

For the IFSn value, we followed the same approach as per the CW; we selected three actions, which are decreasing, maintaining, and increasing the value as $A_3 = \{(\text{IFSn})-1, \text{IFSn}, (\text{IFSn}+1)\}$. The IFSn values per category vary as follows:
\vspace{0.01cm}
\begin{equation}
\text{IFSn}_\text{min}, \text{IFSn}_\text{max} =
\begin{cases}
    1,10 \quad \text{Voice Category} \\ 
    1,20 \quad \text{Video Category}\\
    1,20 \quad \text{HD Map Category}\\
    1,40 \quad \text{Best-effort Category}\\
\end{cases}
\label{eq:values_ifs_min_max}
\end{equation}

Thus, the actions are selected, and the values will start from the number defined previously, which later will be decreased, kept, or increased depending on the agent's actions. Through a thorough analysis of the EDCA mechanism $CW_\text{min}$, $CW_\text{max}$, and IFSn values \cite{iee802_11p_standard} and extensive experimentation, we have determined the above described optimal values per each service. 

For the last action waiting transmission time $wt$, we utilised the same actions described in \cite{our_sojourn_single_agent}, which we have defined as $A_4$.

\begin{equation}
    wt(d) = a \cdot \left(\frac{wt_{max}(d)}{|\mathcal{A}_4|}\right)
    \label{eq:map_actions}
\end{equation}

The equation \eqref{eq:map_actions} consists of the maximum waiting time $wt_\text{max}$ per category, divided by the size of the set of actions $A_4 = \{0,1,2,..7\}$, and multiply the current action $a$. The $wt_\text{max}$ are $0.92$s, $2$s, $2$s, and $8s$ for voice, video, HD map, and best-effort access categories, respectively. These values were derived from a comprehensive analysis and extensive experimentation, as detailed in our previous study \cite{our_sojourn_single_agent}. Notably, the maximum waiting time for HD maps is chosen to be the same as for video in order to ensure fair channel allocation and prevent any potential deterioration in the video service.

\subsection{Reward}
For the rewards, we selected the utility function equation \eqref{eq:utility_function_penalty}, which contains latency and throughput values as part of the optimisation problem. As described, $\mathcal{V}$ represents the penalties and bonuses and is calculated by the equation \eqref{eq:penalty_v}. The threshold values are described in Table \ref{tab:combined_thresholds}.

\begin{table}[h]
\centering
\caption{Thresholds}
\begin{tabular}{|c|c|c|c|c|}
   \hline
   Thresholds & VO & VI & HD Map & BE \\
   \hline
   Data Rate (Mbps) & 0.1 \cite{dataRate_cisco_voice_szigeti2005end} & 1.25 \cite{dataRateVideo_googleYouTube} & 1.25 & 1.0 \\
   \hline
   Delay (ms) & 150 \cite{latency_cisco} & 100 \cite{dataRate_cisco_voice_szigeti2005end} & 100 \cite{5gaa_delay} & 1000 \\
   \hline
\end{tabular}
\label{tab:combined_thresholds}
\end{table}
\begin{algorithm}
\caption{Q-Learning}
\label{algo:q_learning}
\begin{algorithmic}[1]
\STATE Initialize the agent with epsilon $\epsilon
$ , number of actions $k$.
\
\FOR {number\_of\_episodes}
\WHILE{STATUS}
  \STATE get observation of the environment
  \STATE extracts Velocity, Location
  \STATE calculates Sojourn Time
  \STATE provide state \\ $\mathcal{S}_{cw} = \{T_v,C,T_{cv},CW_{min},CW_{max}\}$
  \STATE provide state \\ $\mathcal{S}_{ifs} = \{CW_{min}, CW_{max}, \mathcal{A}_{cw}, IFSn\}$
  \STATE provide state \\ $\mathcal{S}_{wt} = \{S_j,T_v,C,T_{cv}\}$
  \STATE $\text{agent}_{cw}$ select $a_{cw} = choose\_action(\mathcal{S}_{cw},\mathcal{A}_{cw})$ Algorithm (\ref{algo:choose_action})
  \STATE $\text{agent}_{ifs}$ select $a_{ifs} = choose\_action(\mathcal{S}_{ifs},\mathcal{A}_{ifs})$
  \STATE $\text{agent}_{cw}$ select $a_{wt} = choose\_action(\mathcal{S}_{wt},\mathcal{A}_{wt})$
  \STATE agents sent actions to vehicle $v$
  \STATE The AV starts transmission by equation \eqref{eq:constraint_x} $y_c = \mathbb{I}(wt - (t - t_0) = 0)$
  \STATE RSU calculates the $r_t$ according to the Utility function \eqref{eq:utility_function_penalty}
  \STATE agents received next\_observation, $r_t$, STATUS
  \STATE update $Q(s,a)$ per agent by using equation \eqref{eq:q_learning_utility}
\ENDWHILE
\ENDFOR
\STATE \textbf{Finish the subroutine}
\end{algorithmic}
\end{algorithm}

\begin{algorithm}[h]
\caption{choose\_action(S,A)}
\label{algo:choose_action}
\begin{algorithmic}[1]
  \STATE c = extract\_category(S)  
      \IF {$p(\epsilon) < \epsilon$}
        \STATE $a^* = random(\mathcal{A})$ 
      \ELSE
        \STATE $a^* = \arg\max_a Q(s, a)$
      \ENDIF
      \IF {$S_{wt}$}
        \STATE $value = a \cdot w_{max}(d)/|\mathcal{A}|$
      \ELSIF {$S_{ifs}$}
        \STATE $value = (IFSn_t)-1+a$
      \ELSIF {$S_{cw}$}
        \STATE $value = \{\frac{CW_t-1}{2}, CW_t, (CW_{t}*2-1)\}$
      \ENDIF
\RETURN value
\end{algorithmic}
\end{algorithm}

\subsection{Algorithm}
The algorithm for both hierarchical and individual learning agents is described in the Algorithm (\ref{algo:q_learning}). Firstly, the agent receives an observation from the environment. It then extracts the velocity and location to calculate the Sojourn Time, which is crucial for the state space for the agent in charge of finding the suitable waiting transmission time. Then, each agent selects an action, and this action is forwarded to the AV, when received the AV starts the transmission by following equation \eqref{eq:constraint_x} $y_c = \mathbb{I}(wt - (t - t_0) = 0)$. Finally, the Q table values are updated.
\section{Simulation} \label{simulation}
The simulator OMNet++ \cite{omnetppOMNeTDiscrete}, Inet Framework \cite{omnetppINETFramework}, veins \cite{veins}, and SUMO \cite{SUMO} tools have been selected to recreate a vehicular traffic flow scenario that uses the wireless standard IEEE802.11p. OMNet++ and Inet offer the full stack for the Physical and MAC layer for IEEE802.11p. Veins and SUMO provide the traffic flow connectivity to OMNet++. Finally, for implementing the RL algorithm, we utilised veinsgym \cite{veinsgym} developed by M. Schettler with its corresponding adjustment to work with our environment. Please refer to Table \ref{table:simulation_parameters} for a detailed breakdown of our simulation parameters.

Each vehicle enters the scenario every $0.66$s, as per data provided by Traffic and Accident Data Unit / North East Regional Road Safety Resource \cite{dataset}. For our analysis, we selected a rush hour period that contains a $2,376$ number of vehicles. The route or trajectory the AVs follow is assigned randomly. Throughout the simulation analysis, we primarily considered two key performance indicators (KPIs): latency and throughput. Latency is defined as the time difference between the generation and reception of a packet, while throughput measures the total number of received packets within a specific timeframe.
\vspace{0.1cm}
\begin{table}[ht]
\caption{Simulation Parameters}
\begin{center}
\begin{tabular}{|c|c|}
\hline
\textbf{\textit{Parameter}} & \textbf{\textit{Value}} \\
\hline
Tile Dimension & 300x100 meters  \\
\hline
Episodes & 50  \\
\hline
Episode duration & 250s  \\
\hline
$\epsilon-greedy$ & 0.2  \\
\hline
Discount Factor $\gamma $& 0.99  \\
\hline
Learning Rate $\epsilon $& 0.1  \\
\hline
$\alpha_1 $, $\alpha_2 $ & 0.3 and 0.7 respectively  \\
\hline
Simulation time & 250s  \\
\hline
Vehicle Density & varies according traffic flow  \\
\hline
Coverage Area & 200m  \\
\hline
Vehicle max speed & 17m/s  \\
\hline
Vehicle Acceleration  & {2.6 m}/{$s^2$} petrol cars\cite{acceleration_deceleration_petrol_car}. \\
\hline
Vehicle Deceleration  & {4.5 m}/{$s^2$} petrol cars\cite{acceleration_deceleration_petrol_car}.  \\
\hline
Tx Power & 200 mW  \\
\hline
Frequency & 5.9 GHz  \\
\hline
Bandwidth & 10MHz  \\
\hline
Best-effort data rate & 28Mbps  \\
\hline
HD Map data rate & 4Mbps \cite{5gaa_delay} \\
\hline
Video data rate & 5Mbps  \cite{dataRateVideo_googleYouTube} \\
\hline
Voice data rate & 100kbps \cite{dataRate_cisco_voice_szigeti2005end} \\
\hline
TXOP limit & Disabled as per standard  \\
\hline
\end{tabular}
\label{table:simulation_parameters}
\end{center}
\end{table}
\vspace{0.1cm}
\section{Result and Analysis} \label{results}
Results of the proposed RL solution, employing three agents controlling CWmin-max pair, IFSn, and waiting transmission time (wt) values, are compared with several benchmarks. These include the current standard IEEE802.11p and EDCA; the single agent CWmin; the two agents which include former and IFSn agent; eight fixed number of actions \cite{RL_cw_simple} approach, and three actions \cite{q_learning_fairness} approach. The eight-fixed approach represents the fixed contention window values in the set of actions $A_1'$, and the three actions correspond to the set of three actions $A_2$. Thus, we adapted our RL algorithm to function with this set of actions. 
The comparison is presented in three scenarios according to the number of agents involved. Initially, we evaluate the solution with the single agent CWmin-max with a set of actions $A_2$, and values in equation \eqref{eq:values_cw_min_max}, the results are evaluated against IEEE802.1p without EDCA and with EDCA, the eighth fixed set of actions $A_1'$, and three set of actions $A_2$. Subsequently, the comparison extends to a two-agent approach against CWmin-max and IEEE802.1p EDCA. Finally, the performance of the three-agent solution is assessed against both the single-agent and two-agent configurations, as well as IEEE802.11p and EDCA standards (QoS). In the graph results, the legends are defined as follows: Non-Qos refers to the standard IEEE802.11p without EDCA, Qos represents the EDCA, CWminFixed denotes the set of eight actions $A_1'$, CWmin illustrates the set of three actions $A_2$, CWminmaxIFS represents the two-agent approach with both Agent CWmin and IFS, and finally CWminmaxIFS\_STime represents the three-agent solution comprising Agent CWmin, IFS, and wt.
\vspace{0.01cm}
\subsection{Single Agent CWmin-CWmax Pair Values}
Firstly, the simulation is performed with one single agent. The agent for waiting transmission time and IFS are excluded. Thus, only the agent CWminmax acts on the environment.
\vspace{0.01cm}
\subsubsection{Latency}\label{one_agent_results}
\begin{figure}[h]
\centering
\includegraphics[width=\linewidth]{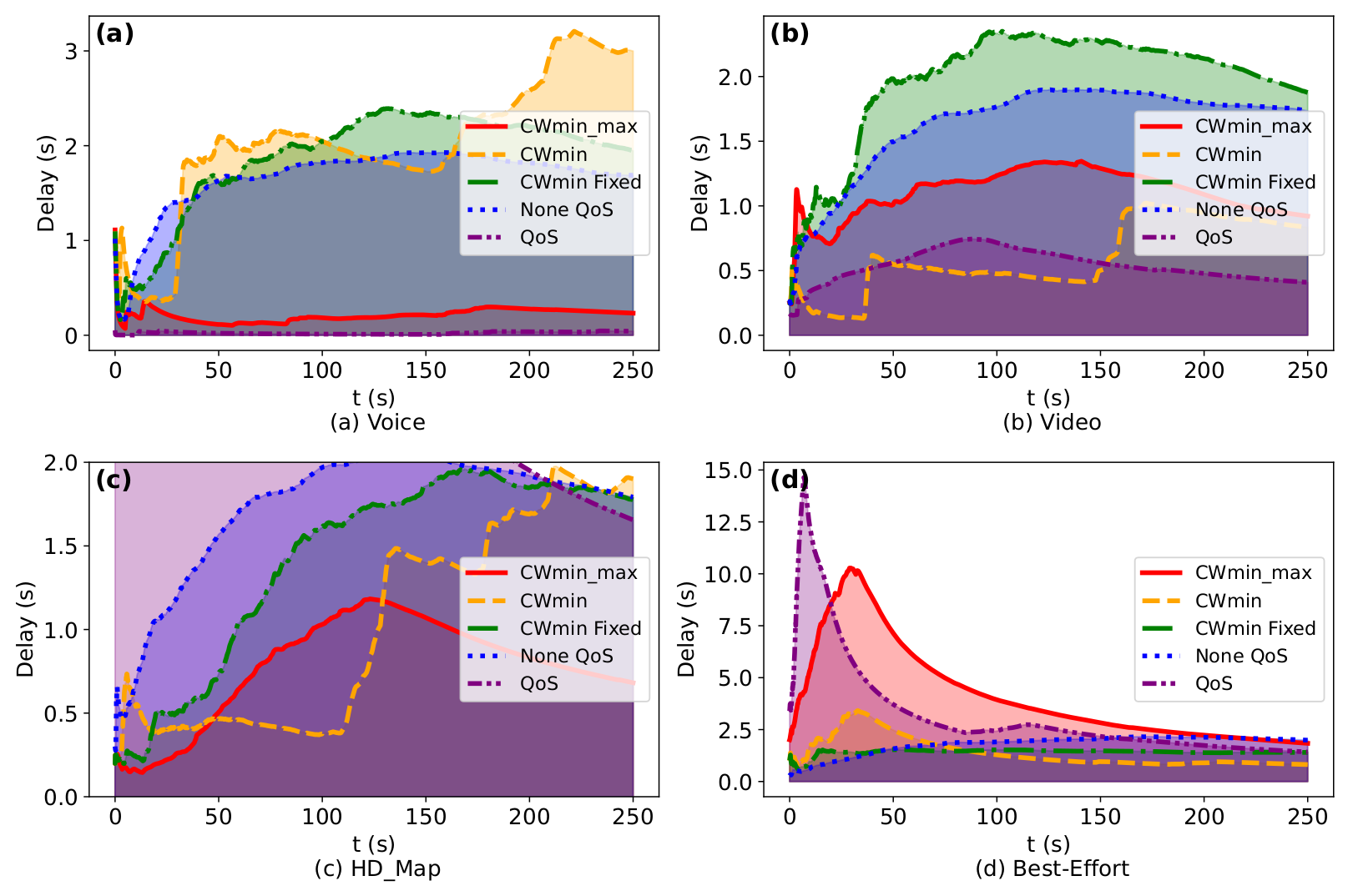}
\caption{Time domain latency comparison between the set of three actions CWmin, the set of eight CWmin Fixed values, None QoS, QoS, and single agent CWminmax. (a) Voice, (b) Video, (c) HD Map, and (d) Best-Effort.}
\label{fig:time_domain_latency_cwminmax}
\end{figure}

\begin{figure}[h]
\centering
\includegraphics[width=\linewidth]{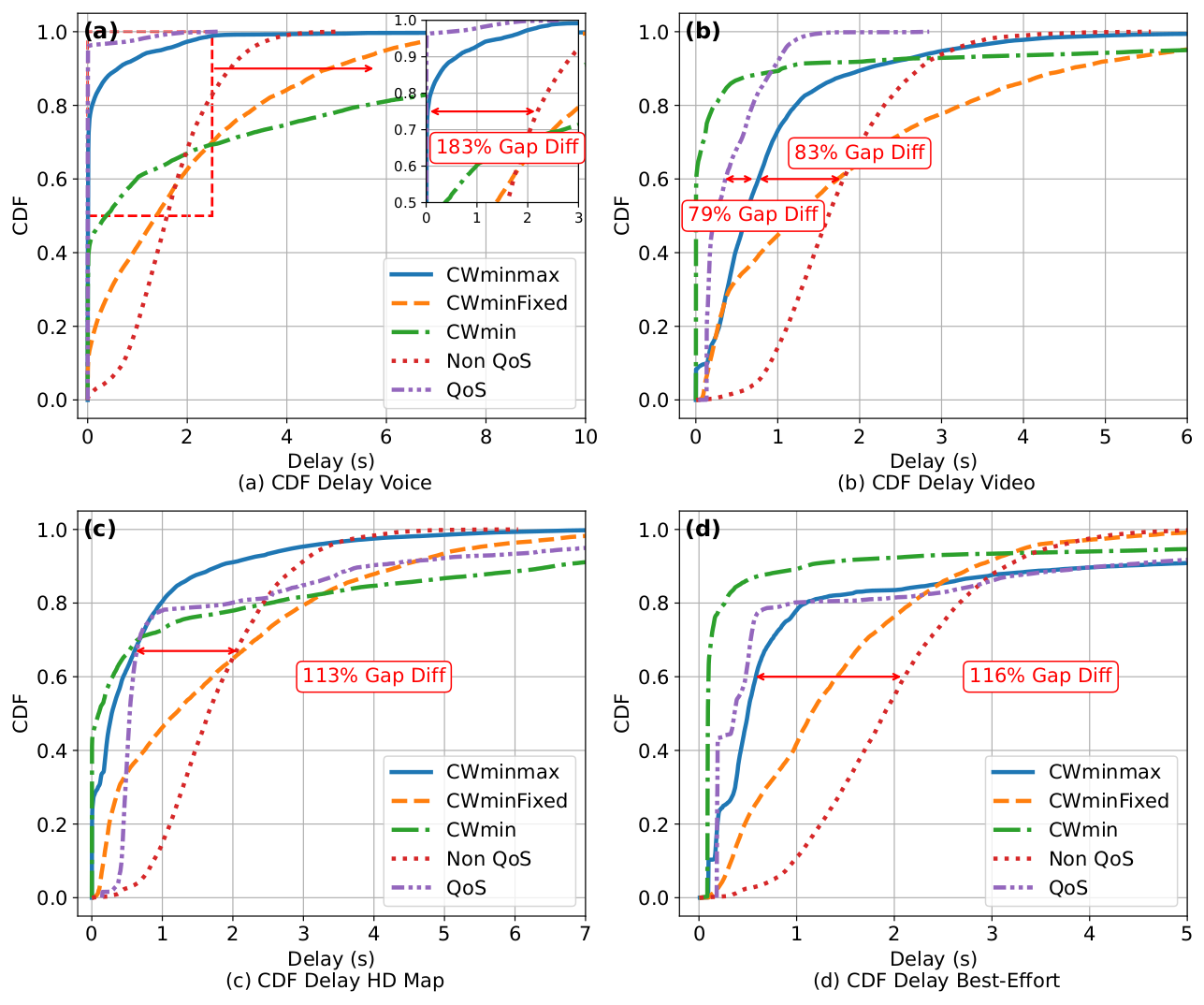}
\caption{CDF latency comparison between the set of three actions CWmin, the set of eight CWmin Fixed values, None QoS, QoS, and single agent CWminmax. (a) Voice, (b) Video, (c) HD Map, and (d) Best-Effort.}
\label{fig:cdf_latency_cwminmax}
\end{figure}

As shown in Fig. \ref{fig:time_domain_latency_cwminmax} (a), it is clear that the CWminmax approach performed with lower latency while comparing voice service results against the fixed set of eight actions (CWmin Fixed, $A_1'$), and the set of three actions (CWmin, $A_2$) methods with a difference of more than $200\%$. However, compared to the EDCA QoS, the CWminmax approach has a percentage difference of $137\%$. The results show voice service is prioritised; though it does not offer the threshold latency of $100$ms, it produced an average latency of $220$ms. For video, as illustrated in Fig. \ref{fig:time_domain_latency_cwminmax} (b), the latency results for CWminmax are not similar or lower than latency compared to the QoS. Nevertheless, the video service is prioritised over best-effort. This outcome is also related to the new HD Map data type prioritization, which indicates the channel must be shared. In Fig. \ref{fig:time_domain_latency_cwminmax} (c), it is observed that HD map latency results for the new approach CWminmax produced lower latency than all the other solutions, which is a desirable result. Considering the results in Fig. \ref{fig:time_domain_latency_cwminmax} (d), the BE latency is higher than the other approaches, which is an expected outcome as it is the lowest priority. The findings indicate that the utilization of CW minimum and CW maximum pair values effectively prioritizes the QoS for HD maps. However, this approach may lead to a slight compromise in the quality of voice and video services.
\vspace{0.01cm}
\vspace{0.001cm}
The Cumulative Distribution Function (CDF) figure shown in \ref{fig:cdf_latency_cwminmax} offers a broader perspective of how the CWminmax approach tries to achieve the same values as the EDCA in the other types of services while providing priority to the new HD map data type. There is a notable improvement compared to the None QoS with a gap difference of $183\%$, $83\%$, $113\%$, and $116\%$ for Voice, Video, HD Map, and BE, respectively. For video, we observed that the QoS performed with lower latency and a gap difference of $79\%$ compared to CWminmax. Nevertheless, the latency values for video service show the prioritization of the service compared with the other approaches, CWmin, CWminFixed, and QoS. The EDCA QoS outperforms the CWminmax method in both voice and video. It is concluded that the single agent CWminmax improved the latency compared to other solutions, showing a focus on fairness. However, it is insufficient to ensure satisfactory QoS for voice and video compared to the standard without EDCA. This is because the standard QoS only considers two data types with higher priorities: voice and video.
\vspace{0.01cm}
\vspace{0.001cm}
\subsubsection{Throughput}
\begin{figure}[h]
\centering
\includegraphics[width=\linewidth]{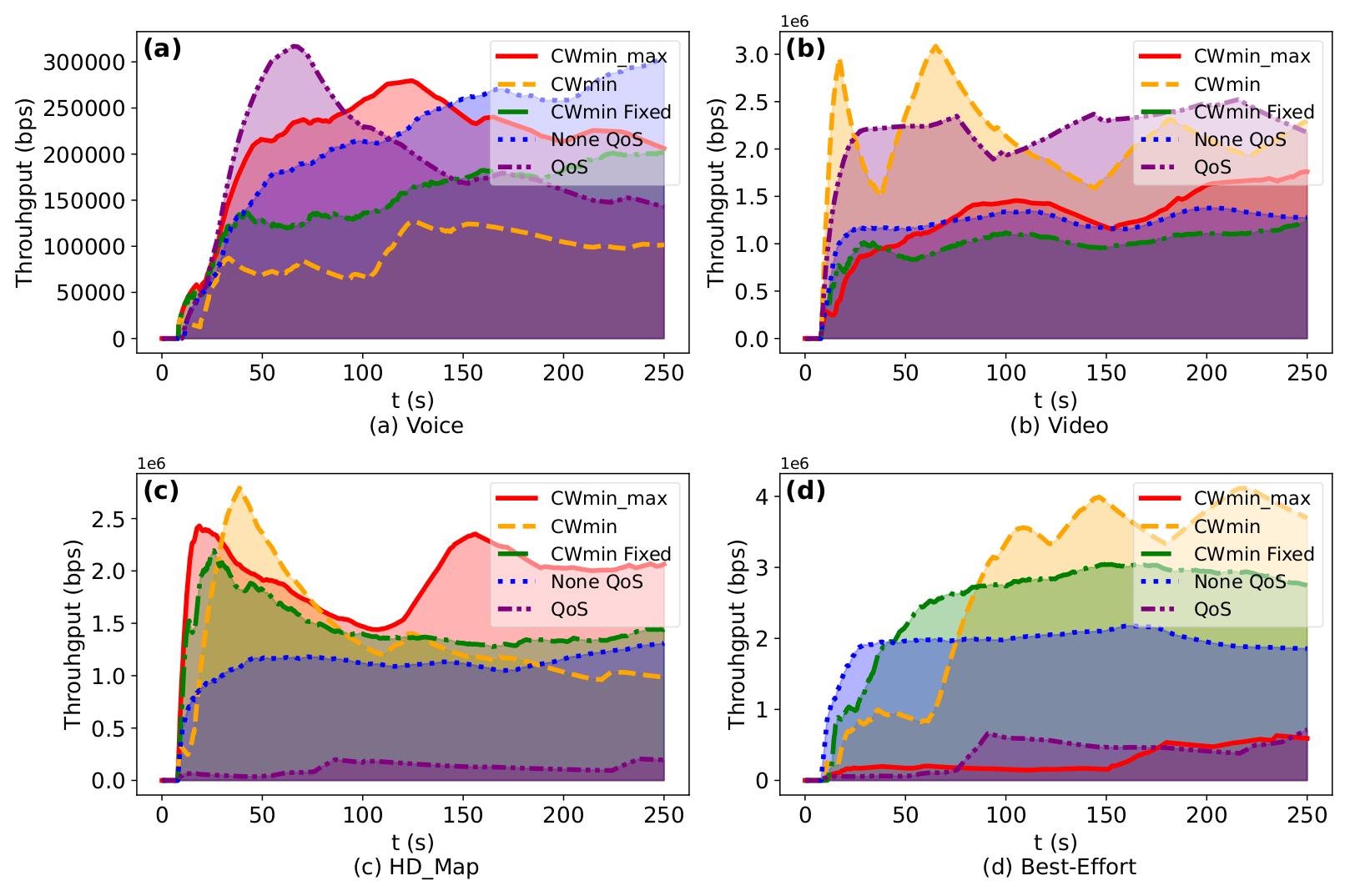}
\caption{Time domain throughput comparison between the set of three actions CWmin, the set of eight CWmin Fixed values, None QoS, QoS, and single agent CWminmax. (a) Voice, (b) Video, (c) HD Map, and (d) Best-Effort.}
\label{fig:time_domain_throughput_cwminmax}
\end{figure}

In Fig.  \ref{fig:time_domain_throughput_cwminmax}, it is seen how the approach of modifying the CWmax also improves the throughput. Starting from the evaluation of voice service, in Fig. \ref{fig:time_domain_throughput_cwminmax} (a), it is observed the CWminmax results displayed a higher stability with a $22\%$ throughput increase over the EDCA (QoS). In comparison, the CWmin approach does not prioritise between services, though it shows fairness between them as the authors intended it in \cite{q_learning_fairness} for the set of eight actions. When the EDCA is not enabled (Non-QoS), the CWmin fixed approach demonstrates similar behaviour in terms of fairness. For the video service, in Fig.  \ref{fig:time_domain_throughput_cwminmax} (b), we could observe a decrease in throughput because we included a new HD Map data type, which indicates the channel must be shared between the services. Nevertheless, the throughput is maintained in the threshold specified in Table \ref{tab:combined_thresholds}. Furthermore, the HD Map throughput displayed in Fig.  \ref{fig:time_domain_throughput_cwminmax} (c) shows an improvement around $100\%$ compared to the EDCA (QoS). Additionally, Fig.  \ref{fig:time_domain_throughput_cwminmax} (d) illustrates that BE maintains a lower priority than the other approaches.

\begin{figure}[h]
\centering
\includegraphics[width=\linewidth]{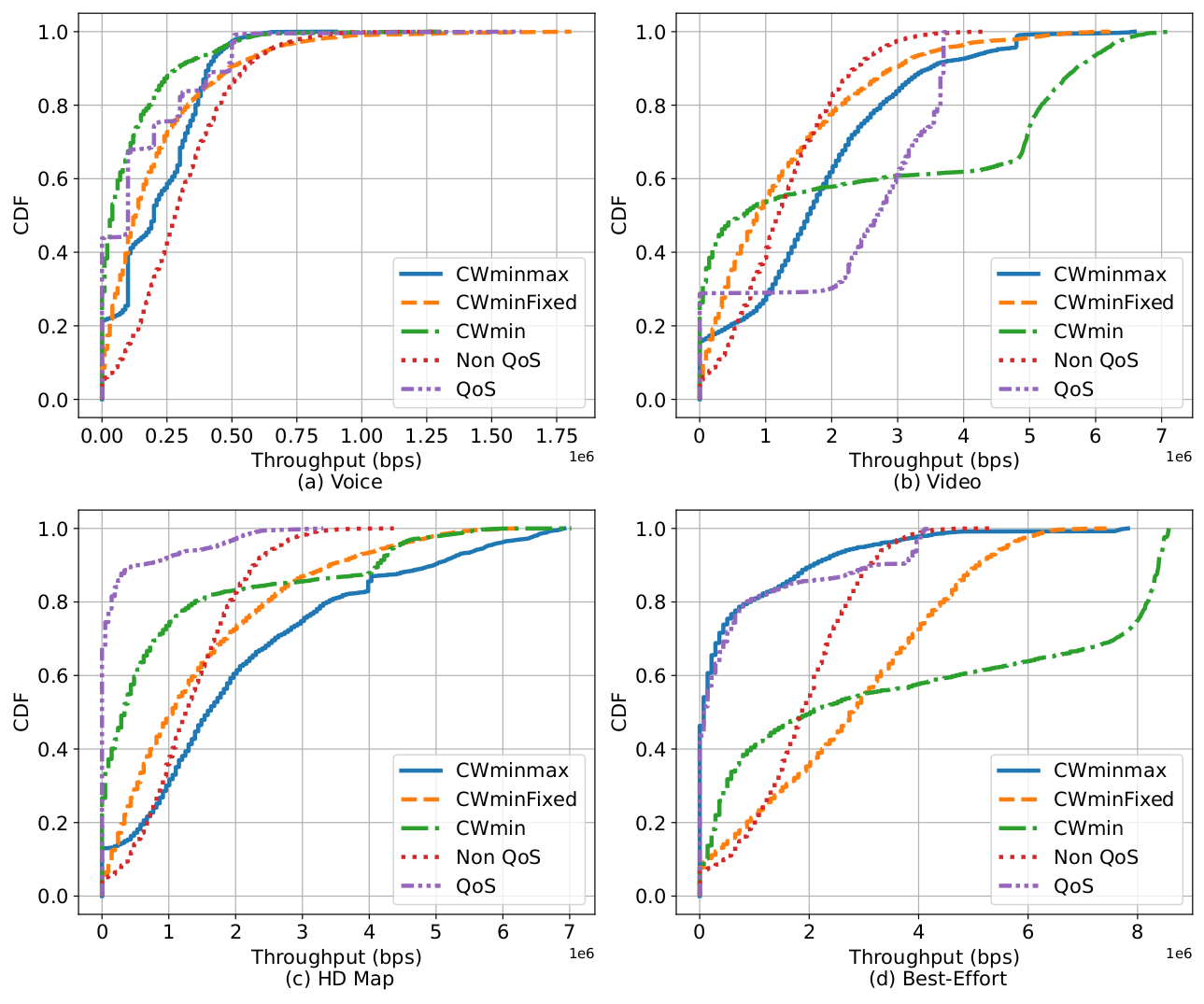}
\caption{CDF throughput comparison between the set of three Actions CWmin, the set of eight CWmin Fixed values, None QoS, QoS, and single agent CWminmax. (a) Voice, (b) Video, (c) HD Map, and (d) Best-Effort.}
\label{fig:cdf_throughput_cwminmax}
\end{figure}

From the other approaches, it is seen in Fig.  \ref{fig:cdf_throughput_cwminmax} that the single agent CWminmax performs well in all services. Nevertheless, in Fig. \ref{fig:cdf_throughput_cwminmax} (b), it is observed that QoS provided a higher throughput. On the contrary, the throughput is improved considerably for HD map service, as seen in Fig \ref{fig:cdf_throughput_cwminmax} (c). This observation demonstrated that our approach considered the priorities much more efficiently than the other solution. Even though the solution improved HD map priority, the other service priorities were compromised.

\vspace{0.01cm}
\subsection{Two Agents: CWmin-CWmax Pair, and IFS}\label{two_agent_results}
Despite the improvements in the previous scenario using a single agent to find CWmin and CWmax, it proves insufficient to surpass the QoS provided by the standard QoS for voice and video while including the new data type HD Map. Nevertheless, it demonstrated that the QoS improved by varying the CWmin and CWmax as a pair. Therefore, we considered the advantage of subdividing tasks into smaller units and assigning a new sub-task to a new agent. The results of this approach are described herein. The simulation is conducted with two agents controlling the CWminmax pair and IFS. Subsequently, the agent responsible for managing waiting transmission time is excluded from our solution. Hence, only two agents are working: Agent CWminmax and IFS.
\vspace{0.01cm}
\subsubsection{Latency}
\begin{figure}[h]
\centering
\includegraphics[width=\linewidth]{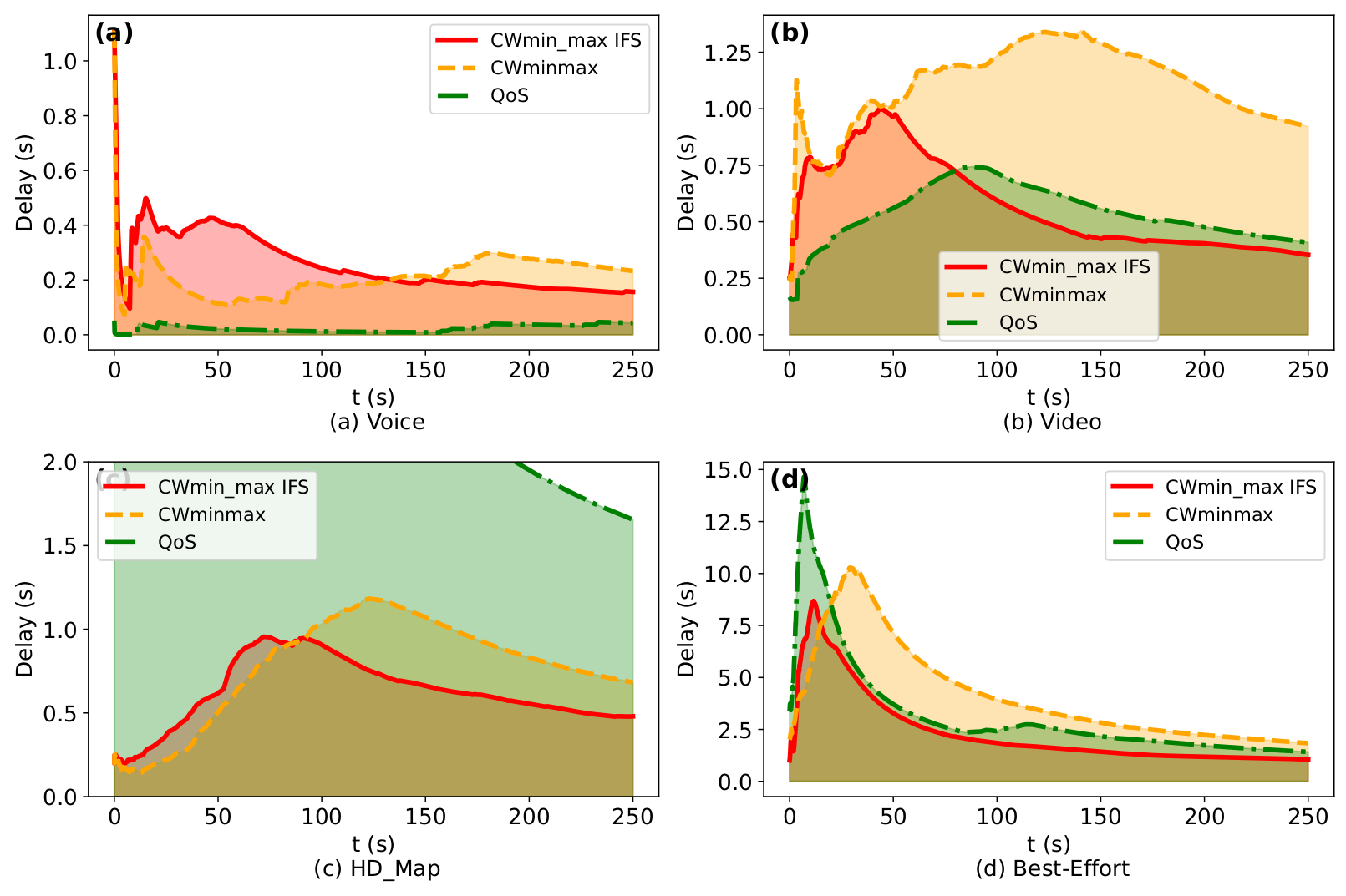}
\caption{Time domain latency comparison between two agents (CWminmax, and IFS), single agent CWminmax, and QoS. (a) Voice, (b) Video, (c) HD Map, and (d) Best-Effort.}
\label{fig:time_domain_latency_IFS_cwminmax}
\end{figure}

With the new method of subdividing the tasks in multi-agents, the latency improved against CWminmax by $61\%$ for video, $29\%$ for HD Map, and $42\%$ for BE as observed in Fig.  \ref{fig:time_domain_latency_IFS_cwminmax} (b), (c), and (d). Compared with the QoS, however, the voice latency average in Fig. \ref{fig:time_domain_latency_IFS_cwminmax} (a) shows higher latency values. For video service, the multi-agent solution produced higher latency at the beginning of the simulation, and later, it followed the QoS.
\vspace{0.01cm}
\vspace{0.01cm}

\begin{figure}[h]
\centering
\includegraphics[width=\linewidth]{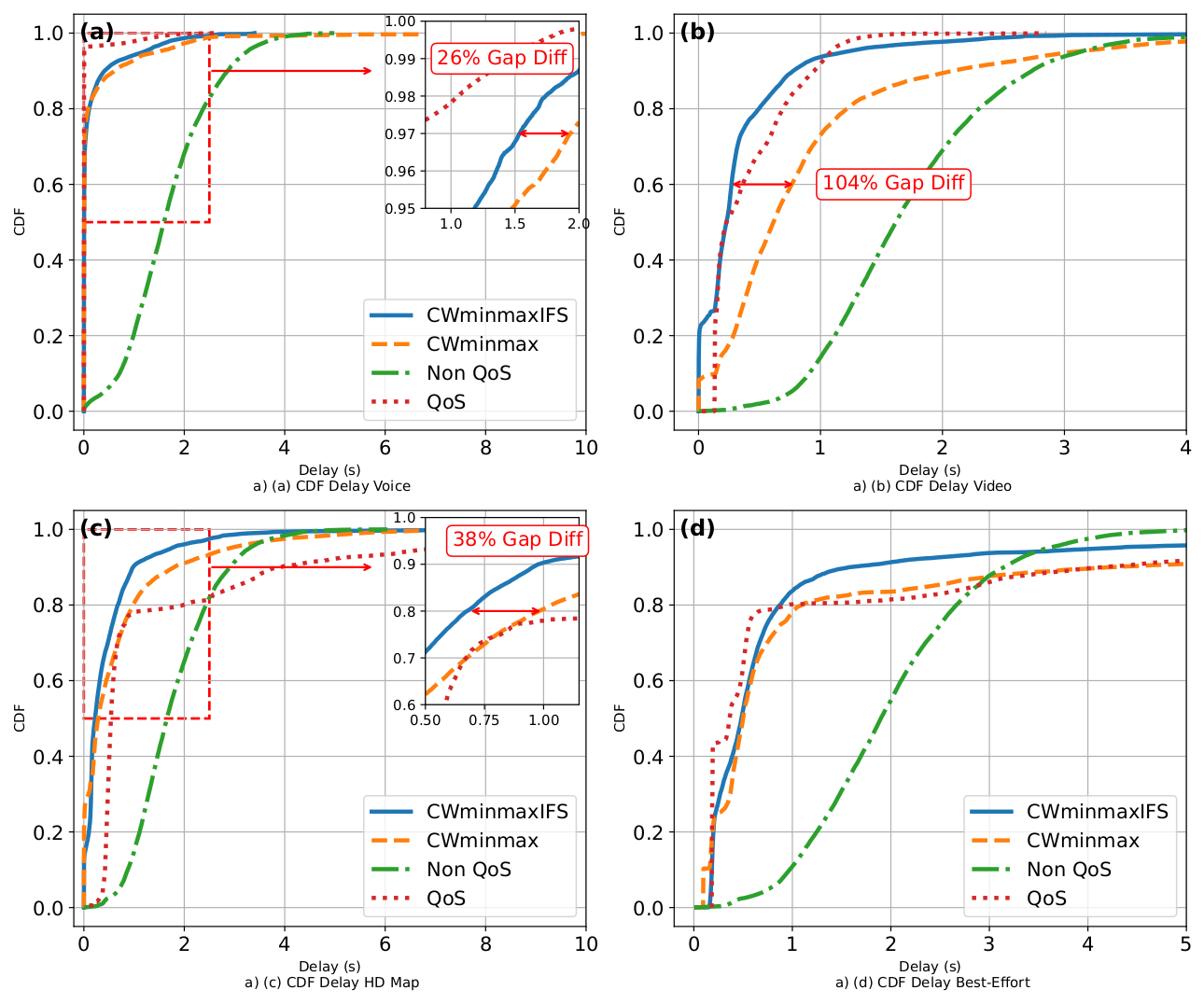}
\caption{CDF latency comparison between, two agents (CWminmax, and IFS), single agent CWminmax, and QoS. (a) Voice, (b) Video, (c) HD Map, and (d) Best-Effort.}
\label{fig:cdf_latency_IFS_cwminmax}
\end{figure}

\vspace{0.01cm}

In the CDF Fig. \ref{fig:cdf_latency_IFS_cwminmax} (a) voice service, from 0.8 to 1.0, the CWminmaxIFS has a difference gap of $26\%$ compared to CWminmax, this indicates that the two agents' solution was able to achieve lower latency in part of the simulation. If we analyze the CDF around the $0.95$ to $0.97$ range, it is observed that CWminmaxIFS latency was $1.5$s, compared to approximately $2.0$s for the CWminmax approach. For the video in Fig.  \ref{fig:cdf_latency_IFS_cwminmax} (b), the line corresponding to CWminmaxIFS is on the leftmost, showing lower latency than the other approaches. It also shows a difference gap of $104\%$ compared to CWminmax. By observing the CDF from $0.0$ to $0.8$, the two-agent approach showed a maximum latency of $0.5$s compared to $1.0$s for CWminmax. For HD Map, there is a difference of $38\%$ against CWminmax. For BE, above $0.8$ of the CDF, there is a higher latency, which shows there is less priority, which is desirable.

\vspace{0.01cm}
\subsubsection{Throughput}
\begin{figure}[h]
\centering
\includegraphics[width=\linewidth]{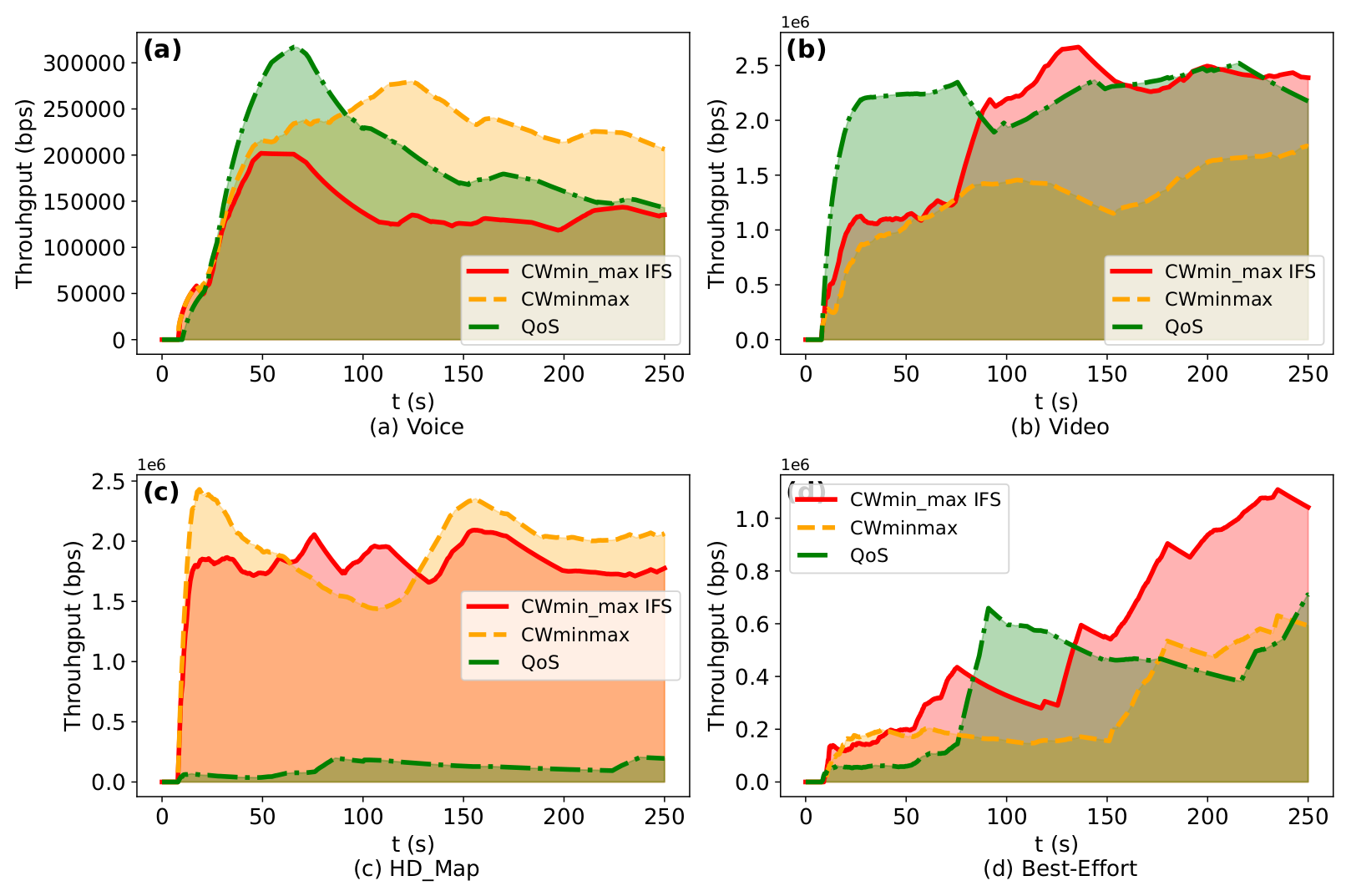}
\caption{Time domain throughput comparison between two agents (CWminmax, and IFS), single agent CWminmax, and QoS. (a) Voice, (b) Video, (c) HD Map, and (d) Best-Effort.}
\label{fig:time_domain_throughput_IFS_cwminmax}
\end{figure}

For the throughput, it is seen in Fig.  \ref{fig:time_domain_throughput_IFS_cwminmax} (a) that voice for the solution CWminmaxIFS has average values closer to the threshold of $100$kbps. For video, as shown in Fig. \ref{fig:time_domain_throughput_IFS_cwminmax} (b), the throughput is higher than the threshold, same as observed in the HD map in Fig. \ref{fig:time_domain_throughput_IFS_cwminmax} (c). Fig. \ref{fig:time_domain_throughput_IFS_cwminmax} (d) displayed the BE latency results. 

\begin{figure}[h]
\centering
\includegraphics[width=\linewidth]{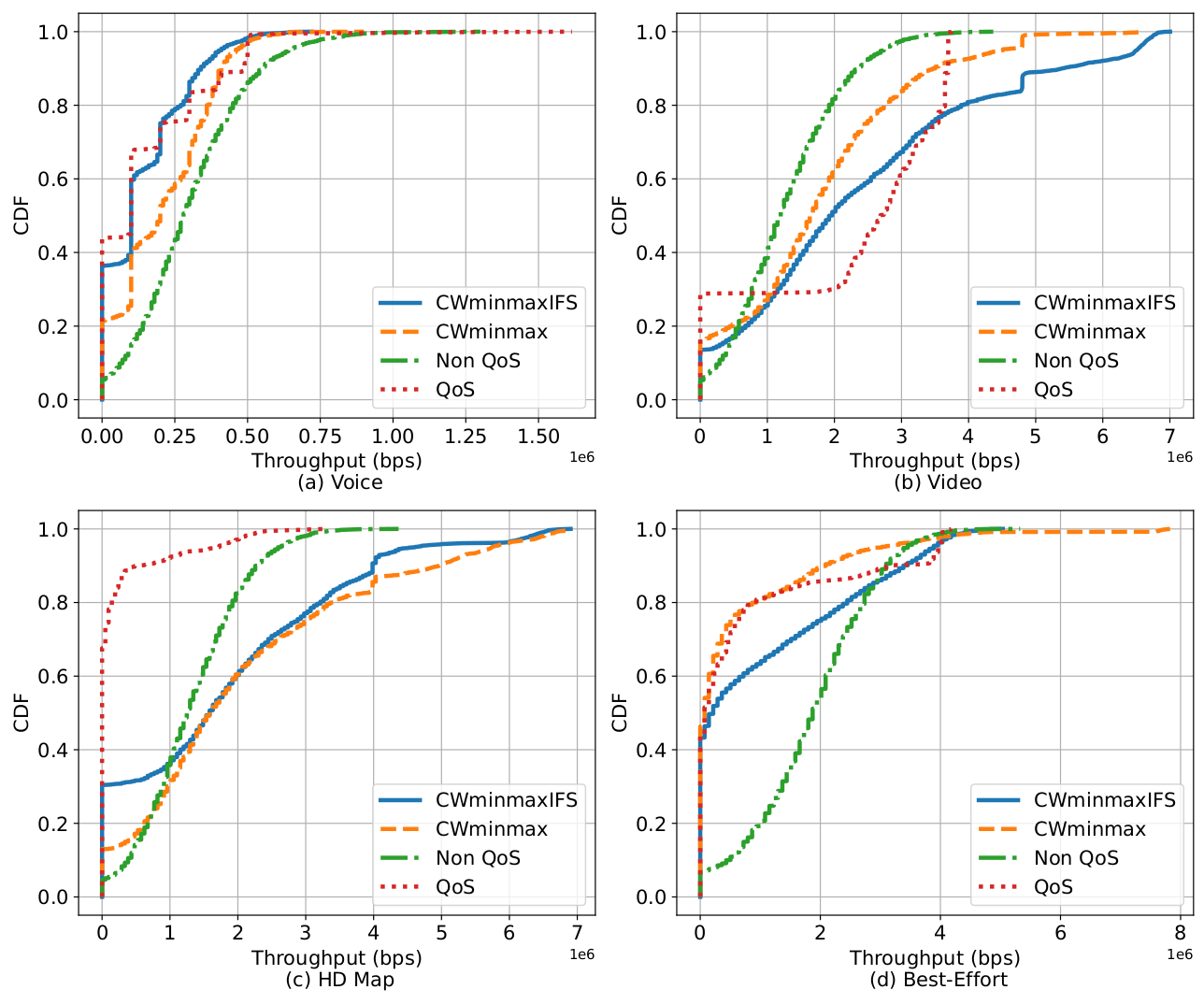}
\caption{CDF throughput comparison between two agents (CWminmax, and IFS), single agent CWminmax, and QoS. (a) Voice, (b) Video, (c) HD Map, and (d) Best-Effort.}
\label{fig:cdf_throughput_IFS_cwminmax}
\end{figure}

The CDF Fig. \ref{fig:cdf_throughput_IFS_cwminmax} (a) shows how close the throughput for the solution of two agents is to the QoS results. For video service in Fig.  \ref{fig:cdf_throughput_IFS_cwminmax} (b) higher values of throughput are displayed. The same is perceived for HD Map in Fig. \ref{fig:cdf_throughput_IFS_cwminmax} (c), and for BE in Fig.  \ref{fig:cdf_throughput_IFS_cwminmax} (d).

\vspace{0.01cm}
\subsection{Three Agents: CWmin-CWmax Pair, IFS, Waiting Transmission Time} \label{three_agent_results}
In the previous scenario, two agents improved the network performance for video compared to one single agent, CWminmax. Nevertheless, there is still room for improvement regarding voice and video. Thus, the three agents, CWminmax pair, IFS, and waiting time, are implemented to increase the network performance.
\vspace{0.01cm}
\subsubsection{Latency}
\begin{figure}[h]
\centering
\includegraphics[width=\linewidth]{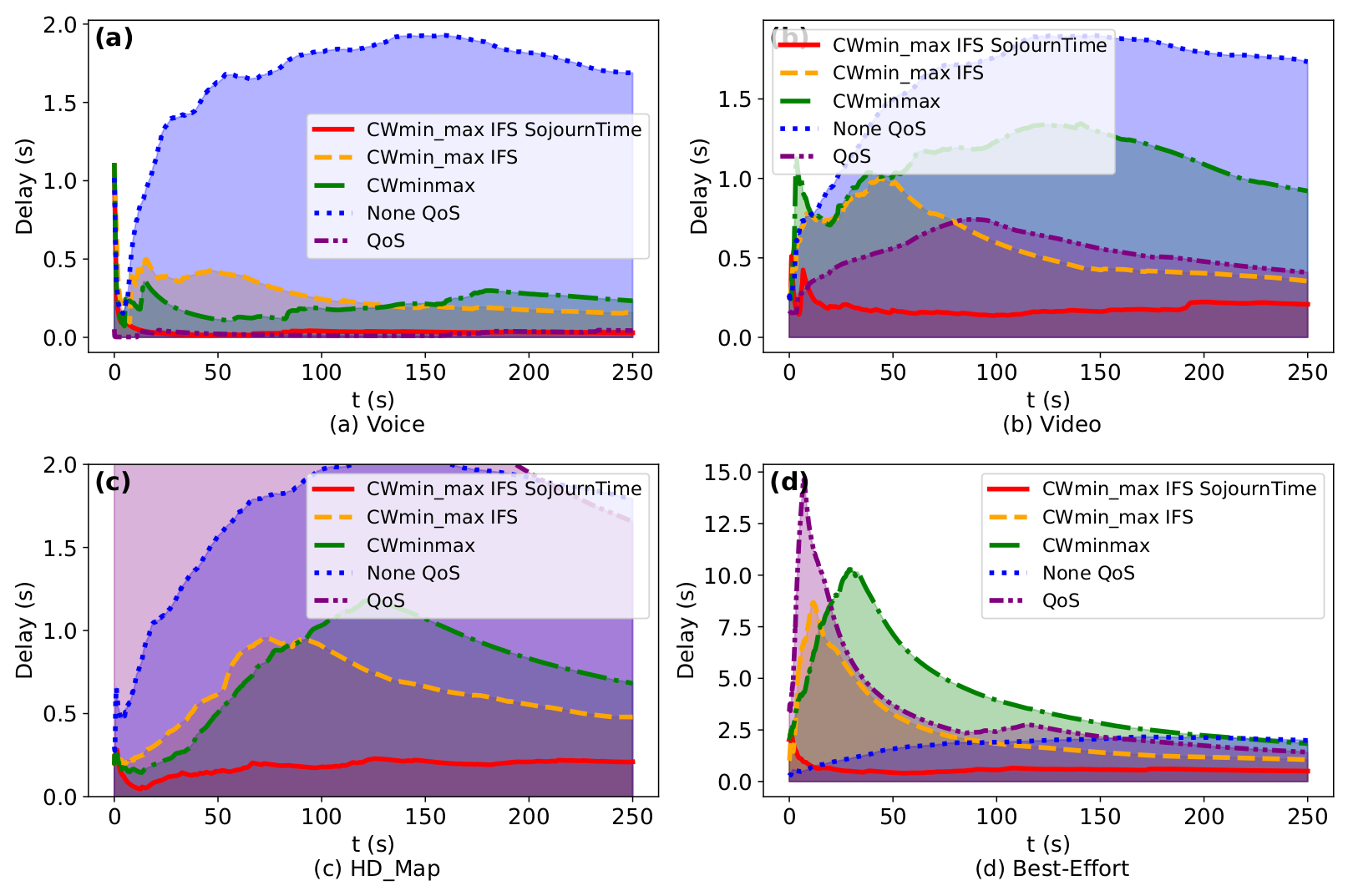}
\caption{Time domain latency comparison between three agents (CWminmax, and IFS, waiting transmission time), two agents (CWminmax, and IFS), single agent CWminmax, None QoS, and QoS. (a) Voice, (b) Video, (c) HD Map, and (d) Best-Effort.}
\label{fig:time_domain_latency_IFS_cwminmax_sojourn}
\end{figure}

As observed in Fig.  \ref{fig:time_domain_latency_IFS_cwminmax_sojourn} (a), the solution integrating the hierarchical and IL agents and standard QoS presented the lowest average latency for voice. The average for the three agents was $0.0293$s, and for the QoS the average was $0.043$s. This comparison with the standard QoS clearly demonstrates the benefits of the three-agent approach, which improves the latency by $31.7\%$. This comparison also underscores the importance of subdividing the tasks to achieve a higher utility. In comparison with the two agents CWminmaxIFS and one agent CWminmax there is an improvement of $81\%$ and $87\%$, respectively. 
For video service in Fig.  \ref{fig:time_domain_latency_IFS_cwminmax_sojourn} (b), we also see an improvement of $26\%$, $41\%$, and $49\%$ compared to CWminmax, CWminmaxIFS, and QoS respectively. The improvement is also notable for HD Map Fig.  \ref{fig:time_domain_latency_IFS_cwminmax_sojourn} (c) compared with CWminmax, CWminmaxIFS, and QoS, the latency decreases by $69\%$, $52\%$, and $87.3\%$, respectively. The improvement is also observed for BE Fig.  \ref{fig:time_domain_latency_IFS_cwminmax_sojourn} (d). Compared with CWminmax, a decrease of $72\%$ is perceived, with CWminmaxIFS a reduction of $52\%$, and with QoS, the decrease is $64\%$.

\begin{figure}[h]
\centering
\includegraphics[width=\linewidth]{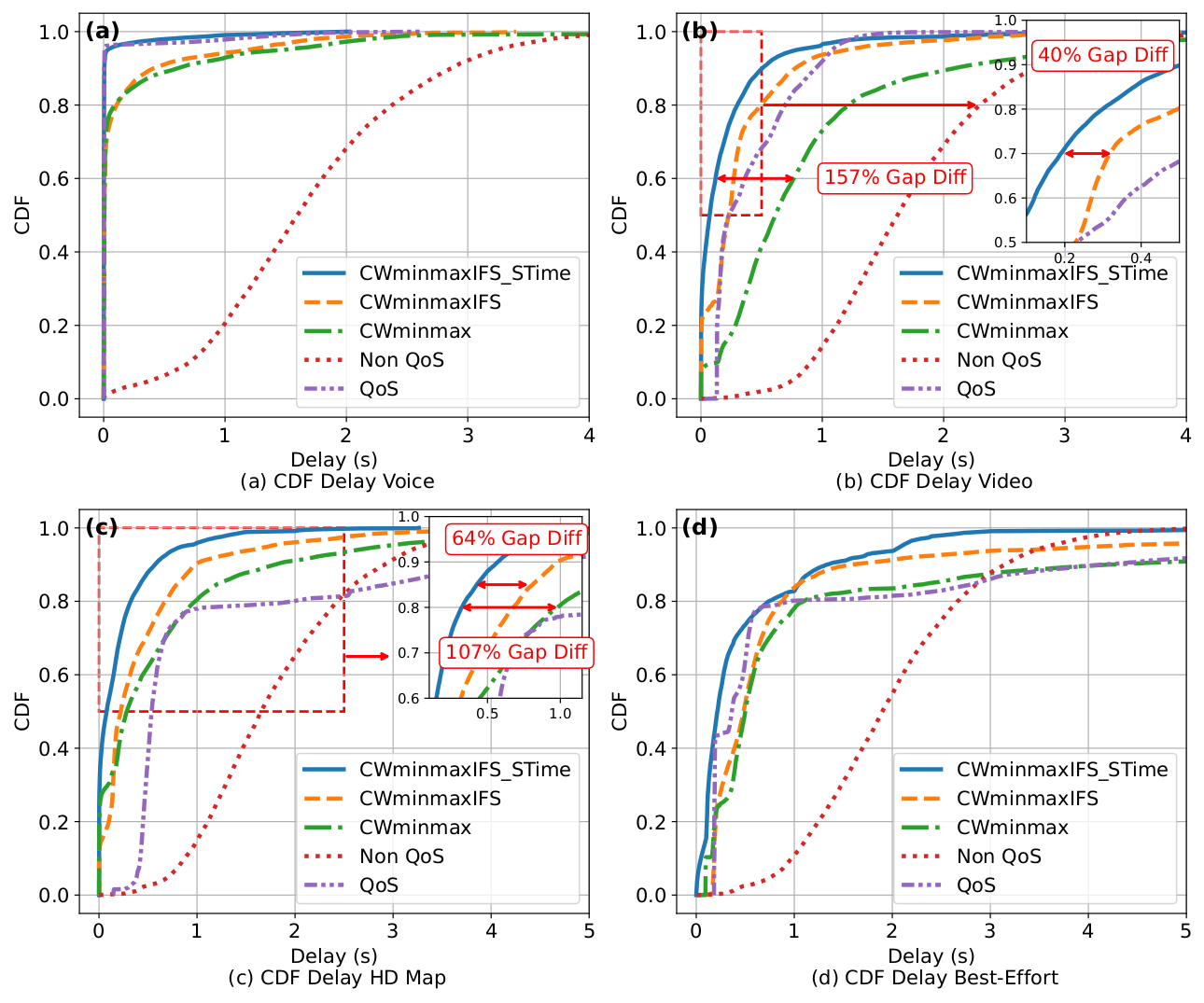}%
\caption{CDF latency comparison between three agents (CWminmax, IFS, waiting transmission time), two agents (CWminmax, and IFS), single agent CWminmax, None QoS, and QoS. (a) Voice, (b) Video, (c) HD Map, and (d) Best-Effort.}
\label{fig:cdf_latency_IFS_cwminmax_sojourntime}
\end{figure}

The CDF demonstrates the same behaviour as observed in the time domain. For voice Fig.  \ref{fig:cdf_latency_IFS_cwminmax_sojourntime} (a), the lines for QoS and three agents solution almost overlap for voice service. For video Fig.  \ref{fig:cdf_latency_IFS_cwminmax_sojourntime} (b), there is a gap difference of $40\%$ compared to the previous solution CWminmaxIFS, and for HD Map Fig.  \ref{fig:cdf_latency_IFS_cwminmax_sojourntime} (c) the difference is $64\%$. This improvement indicates that the new approach could guarantee the QoS for all the services even with the addition of a new data type, HD Map.

\subsubsection{Throughput}
\begin{figure}[h]
\centering
\includegraphics[width=\linewidth]{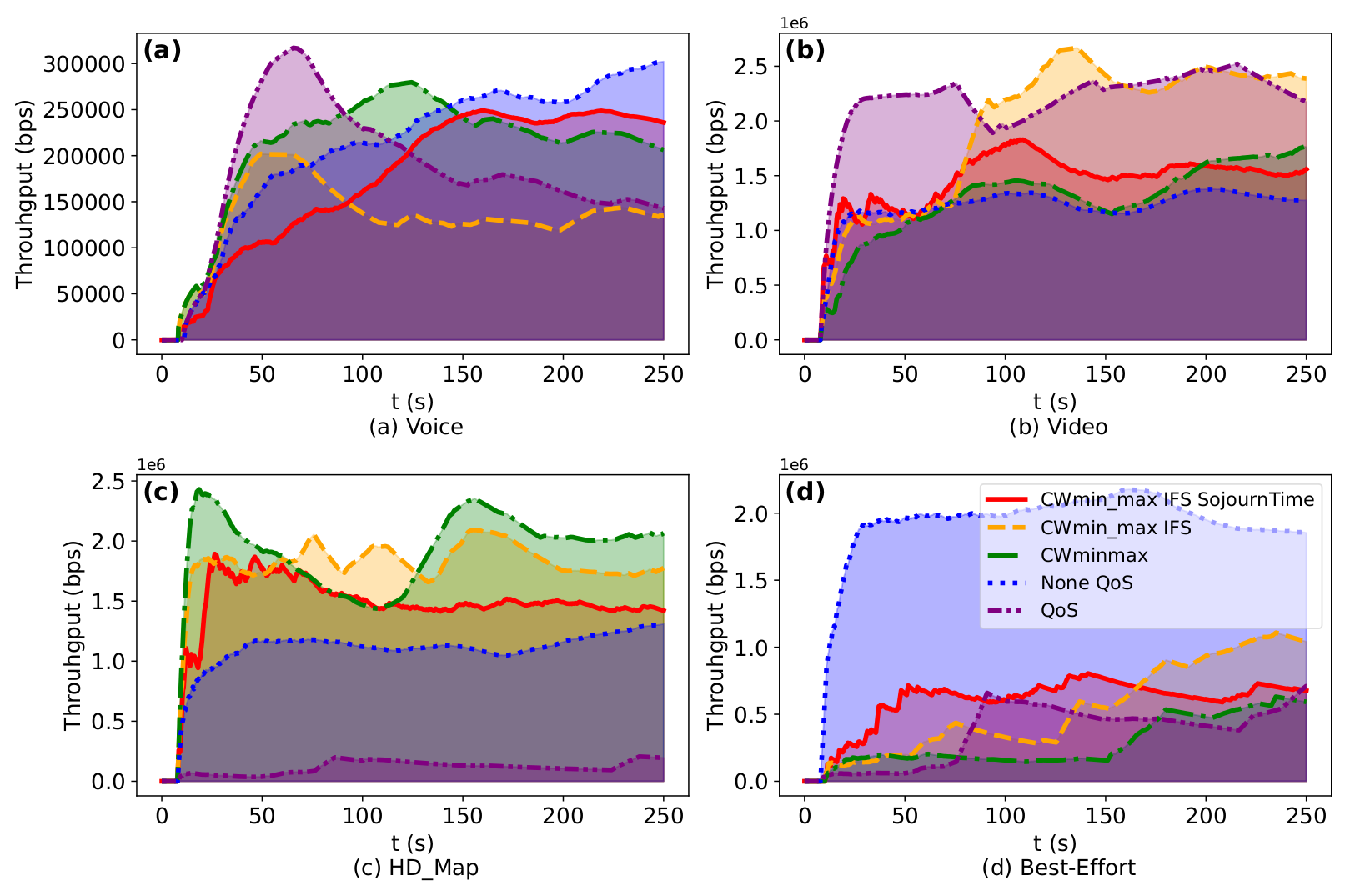}
\caption{Time domain throughput comparison between three agents (CWminmax, and IFS, waiting transmission time), two agents (CWminmax, and IFS), single agent CWminmax, None QoS, and QoS. (a) Voice, (b) Video, (c) HD Map, and (d) Best-Effort.}
\label{fig:time_domain_throughput_IFS_cwminmax_sojourn}
\end{figure}

From Fig.  \ref{fig:time_domain_throughput_IFS_cwminmax_sojourn} (a), a slight improvement in throughput for voice is observed. For video and HD maps in Fig. \ref{fig:time_domain_throughput_IFS_cwminmax_sojourn} (b) and (c), the solution of three agents was able to keep the value close to the threshold. For BE displayed in Fig. \ref{fig:time_domain_throughput_IFS_cwminmax_sojourn} (d), the throughput tries to keep the values lower than the threshold. 

\begin{figure}[h]
\centering
\includegraphics[width=\linewidth]{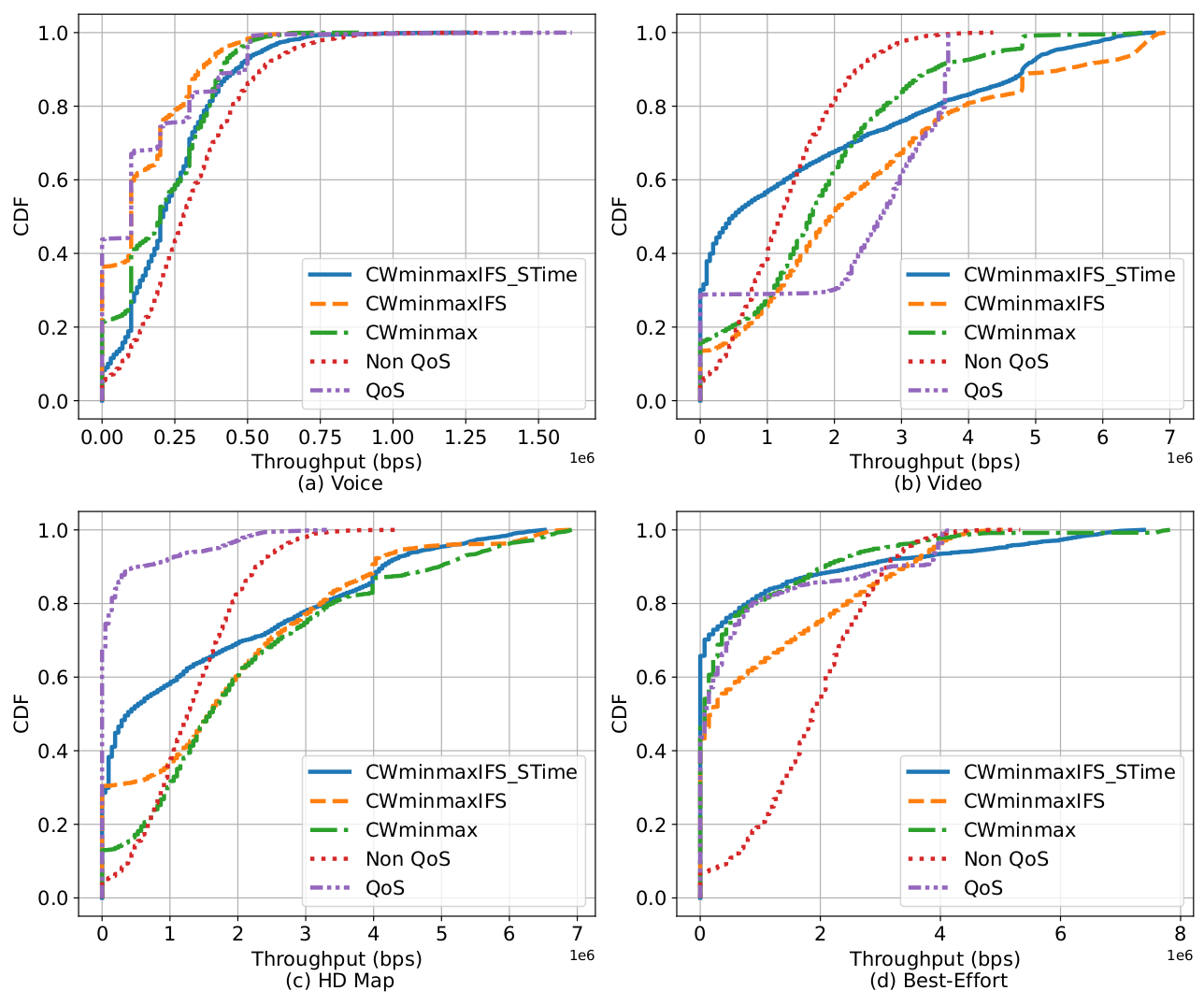}
\caption{CDF throughput comparison between three agents (CWminmax, and IFS, waiting transmission time), two agents (CWminmax, and IFS), single agent CWminmax, None QoS, and QoS. (a) Voice, (b) Video, (c) HD Map, and (d) Best-Effort.}
\label{fig:cdf_throughput_IFS_cwminmax_sojourn}
\end{figure}

In the CDF in Fig. \ref{fig:cdf_throughput_IFS_cwminmax_sojourn}, we could observe the same behaviour as described in the results above regarding Fig. \ref{fig:time_domain_throughput_IFS_cwminmax_sojourn}. Nevertheless, there are some points to notice. For voice service in Fig. \ref{fig:cdf_throughput_IFS_cwminmax_sojourn} (a), the CWminmaxIFS\_STime line corresponding to the three agents displayed higher throughput than the standard Qos. In regards to the video service Fig. \ref{fig:cdf_throughput_IFS_cwminmax_sojourn} (b) illustrated that the throughput is close to the threshold set on Table \ref{tab:combined_thresholds}. For HD maps, throughput results displayed in Fig. \ref{fig:cdf_throughput_IFS_cwminmax_sojourn} (c) reveal that the three agent solution has a similar behaviour as the other solutions, with the difference that the three agent solutions maintain the throughput stable during the simulation. Finally, in Fig. \ref{fig:cdf_throughput_IFS_cwminmax_sojourn} (d), the results for BE presented a lower throughput as expected due to the lowest priority.

\section{Discussion} \label{discussion}
We developed a multi-agent system capable of handling multi-tasks and a multi-service environment to disseminate HD maps for AVs. Our three-agent RL design has been successfully implemented and surpasses the benchmark and other approaches compared with the literature review that aims to mitigate the packet collision problem in IEEE802.11 due to the fixed CW as highlighted in \cite{ieee80211_cw}, and \cite{ieee80211_cw2}. We compared with the benchmark standard IEEE802.11p with and without EDCA, a set of eight actions $A_1'$ studied in \cite{RL_cw_simple} and labelled CWminFixed, and a set of three actions $A_2$ \cite{q_learning_fairness}, labelled CWmin. A distinctive feature of our approach, which sets it apart from other works, is that, unlike others, we assess the system in a multi-service environment. Consequently, we have considered their action space in our RL algorithm to compare our task subdivision with a multi-agent solution in a multi-service environment. After the simulation and result analysis, we could summarize that the above approaches produce fairness between the AV; however, it does not provide any priority for different applications. As suggested by one of the authors in \cite{q_learning_fairness}, future work should be done combining Q\_learning, CW and the EDCA priorities for different types of traffic.
Consequently, we utilised more than just the EDCA and CW; we also considered the IFS. The three-agent solution outperformed the single-agent, two-agent, Qos, Non-QoS, CWmin, CWminFixed, and CWminmax solutions.

\section{Conclusion} \label{conclusion}
This paper has demonstrated the advantage of subdividing the task into subtasks with a multi-agent approach. The methodology consists of a cross-layer design that permits the MAC layer to access the suggested values directly from the application layer. Furthermore, the design emphasises hierarchical and independent learning machine learning (ML) techniques. Specifically, two agents operate within a hierarchical approach, while another agent works independently. By implementing this strategy, the three-agent solution has improved latency to efficiently accommodate the wireless resource for each AV to disseminate HD map updates while simultaneously interacting with different services. Regarding priority for providing QoS, our solution outperformed the two approaches, the set of eight fixed contention windows action and another that utilises a set of three actions that increase, maintain, or decrease the CW while accommodating multiple services. Moreover, it surpassed the standard IEEE802.11p EDCA by  $31\%$, $49\%$, $87.3\%$, and $64\%$ for Voice, Video, HD Map, and Best-effort respectively.

\section*{Acknowledgment}
This work was partly funded by EPSRC with RC Grant reference EP/Y028023/1, UKRI under grant number EP/Y028023/1, the European Horizon2020 MSCA programme under grant agreement  No. 101086228.

\bibliographystyle{IEEEtran}
\bibliography{IEEEabrv,Bibliography}

\end{document}